\documentclass[onecolumn,nobibnotes,nofootinbib,superscriptaddress]{revtex4}
\usepackage{amsmath,amssymb,bm}
\usepackage[a4paper,bindingoffset=0.2in,left=0.8in,right=0.8in,top=1in,bottom=1in,footskip=.25in]{geometry}
\usepackage{graphicx,subfigure,epsfig}
\usepackage{bigints}
\usepackage{color}
\usepackage[colorlinks=true,linkcolor=blue,anchorcolor=blue,citecolor=blue,urlcolor=blue]{hyperref}

\newcommand{\be}{\begin{equation}}
\newcommand{\ee}{\end{equation}}
\newcommand{\bea}{\begin{eqnarray}}
\newcommand{\eea}{\end{eqnarray}}

\newcommand{\nn}{\nonumber\\}
\newcommand{\dif}{{\rm d}}
\newcommand{\del}{\partial}
\newcommand{\lt}{\left}
\newcommand{\rt}{\right}
\newcommand{\bk}{\bm{k}}

\newcommand{\bx}{\bm{x}}
\newcommand{\by}{\bm{y}}
\newcommand{\bu}{\bm{u}}
\newcommand{\bv}{\bm{v}}
\newcommand{\bz}{\bm{z}}
\newcommand{\br}{\bm{r}}
\newcommand{\abar}{\bar{\alpha}_s}
\newcommand{\af}{\alpha_s}
\newcommand{\minus}{\!-\!}

\begin{document}

\title{Extended collinearly-improved Balitsky-Kovchegov evolution equation in target rapidity}
\author{Wenchang Xiang}
\email{wxiangphy@gmail.com}
\affiliation{Guizhou Key Laboratory in Physics and Related Areas, and Guizhou Key Laboratory of Big Data Statistic Analysis, Guizhou University of Finance and Economics, Guiyang 550025, China}
\affiliation{Department of Physics, Guizhou University, Guiyang 550025, China}
\author{Yanbing Cai}
\email{myparticle@163.com}
\affiliation{Guizhou Key Laboratory in Physics and Related Areas, and Guizhou Key Laboratory of Big Data Statistic Analysis, Guizhou University of Finance and Economics, Guiyang 550025, China}
\author{Mengliang Wang}
\email{mengliang.wang@mail.gufe.edu.cn}
\affiliation{Guizhou Key Laboratory in Physics and Related Areas, and Guizhou Key Laboratory of Big Data Statistic Analysis, Guizhou University of Finance and Economics, Guiyang 550025, China}
\author{Daicui Zhou}
\email{dczhou@mail.ccnu.edu.cn}
\affiliation{Key Laboratory of Quark and Lepton Physics (MOE), and Institute of Particle Physics, Central China Normal University, Wuhan 430079, China}


\begin{abstract}
An extended collinearly-improved Balitsky-Kovchegov evolution equation in the target rapidity representation is derived by including the running coupling corrections during the expansion of the ``real" $S$-matrix. We find that the running coupling brings important corrections to the evolution equation, as one can see that there are extra contributions to the evolution kernel once the running coupling is included. To identify the significance of the corrections, we numerically solve the evolution equation with and without the running coupling contributions during the $S$-matrix expansion. The numerical results show that the scattering amplitude is largely suppressed by the running coupling corrections, which indicate that one needs to consider the running coupling contributions during the derivation of the non-linear evolution equation in the target rapidity representation.
\end{abstract}

\maketitle


\section{Introduction}
\label{sec:intro}
The Color Glass Condensate (CGC) effective theory has been approved to be as a powerful theory to describe the strong interactions associated with high energy and density environments. The leading order (LO) CGC calculations which refer to the derivation of the non-linear Balitsky-JIMWLK\footnote{The JIMWLK is the abbreviation of Jalilian-Marian, Iancu, McLerran, Weigert, Leonidov, Kovner.}\cite{B,JIMWLK1,JIMWLK2,JIMWLK3,JIMWLK4} equation and its mean field version known as the Balitsky-Kovchegov (BK) equation\cite{B,K}, have been able to qualitatively describe many phenomenological results, such as the reduced cross-section in deep inelastic scattering (DIS)\cite{IIM,Xiang07}, and single and double inclusive particle production in high energy heavy ion collisions\cite{Levin01,Dumitru05,Levin10,CX,LM13}. However, it has been found that the LO CGC theory is insufficient for direct applications to the phenomenology, since the evolution speed of the scattering amplitude resulting from the LO BK equation is too fast to quantitatively describe the experimental data\cite{JS,JM}.

There are tremendous developments in the calculations of the next-to-leading order (NLO) corrections to the JIMWLK and BK equations in the literature over the past fifteen years\cite{Bnlo,KW,BC08,NLOJIMWLK1,NLOJIMWLK2,NLOJIMWLK3,Zhou19}. The pioneer work towards the NLO corrections to the BK equation was performed by including the running coupling effect in Refs.\cite{Bnlo} and \cite{KW}. A running coupling Balitsky-Kovchegov (rcBK) equation was obtained. It was shown that the growth of the dipole-hadron scattering amplitude resulting from the rcBK equation is significantly slowed down as compared to the LO one. Furthermore, it was found that the rcBK equation gives a rather successful description of the HERA data\cite{JS}. However, the running coupling effect is not the only large higher order corrections to the LO BK or JIMWLK equations. Except the running coupling, the authors in Ref.\cite{BC08} derived the full NLO BK evolution equation by including the quark loops, gluon loops, as well as the tree gluon diagrams with quadratic and cubic non-linearities, they obtained other contributions which are enhanced by double transverse logarithms. Unfortunately, the numerical study of the full NLO BK equation found that the equation is unstable, since the scattering amplitude can decrease as rapidity increasing and can even turn to negative value for small dipoles\cite{LM15,LM16}. The instability was traced back to the radiative corrections enhanced by double transverse logarithms.

To cure the instability problem, one has to resum the radiative corrections enhanced by double transverse logarithms to all orders. Two methods related to this specific issue were proposed\cite{Beuf14,IMMST}, which use different recipes to impose a kinematical constraint to the successive gluon emissions during the rapidity evolution, (i) the kinematical constrain introduced in Ref.\cite{Beuf14} leads to an evolution equation which is similar to the LO BK equation, but is non-local in rapidity $Y$; (ii) the resummation of the leading double logarithms has been performed in the evolution kernel in Ref.\cite{IMMST} resulting in a collinearly-improved Balitsky-Kovchegov (ciBK) equation with a modified kernel, which is still local in rapidity $Y$. It has been shown that these two methods are equivalent in the resummation of the leading double transverse logarithms. It is known that both of them lead to stable evolution equation and give good fits to the small-$x$ HERA data\cite{IMMST2,Cai20,Xiang21}. However, the authors in Ref.\cite{DIMST19} found there are some inconsistencies with the original analysis in Refs\cite{Beuf14,IMMST}. They found that the instability of the full NLO BK is caused by the wrong choice of the rapidity variable which plays the role of the evolution time. Moreover, it has been found that the growth of the saturation exponent with the running coupling constant in the asymptotic region of rapidity has a strong scheme dependence on the resummation method, which should not occur in practice.

It is improtant to point out that the rapidity generally used in all above mentioned BK equations is that of the projectile rapidity $Y$. In terms of the previous experience with the NLO BFKL\footnote{The BFKL is the abbreviation of Balitsky-Fadin-Kuraev-Lipatov.} equation\cite{BFKL1,BFKL2,BFKL3}, where a similar issues were met and eventually cured, the evolution variable should be the rapidity of the hadronic target, $\eta$, rather than the rapidity of the projectile. Inspired by the experience on handling instability problem in NLO BFKL equation, a novel method was proposed to derive the collinearly-improved BK equation in $\eta$-representation (ciBK-$\eta$) from the corresponding evolution equation in $Y$-representation by the change of variable $\eta=Y-\rho$, where $\rho$ is defined as $\rho=\ln(Q^2/Q_0^2)$ with $Q^2$ and $Q_0^2$ to be the hard scale of the projectile and soft scale of the target, respectively\cite{DIMST19}. As a consequence, the evolution variable in the ciBK-$\eta$ equation is the physical rapidity $\eta=\ln(1/x)$, not the rapidity of the dipole projectile, and the ciBK-$\eta$ equation shows a little scheme dependence on the resummation prescriptions.
Although the ciBK-$\eta$ equation has a significant advantage over the relevant one in $Y$-representation, it has recently been found that the evolution equation in $\eta$ or $Y$-representation gives a very similar description of the HERA data\cite{Beuf20}. The reason for this unexpected result could be attributed to several factors: (i) only the dominant part of the ciBK-$\eta$ equation, which is called ``canonical" Balitsky-Kovchegov equation (caBK-$\eta$), was used to study the data in order to avoid the cumbersomely numerical calculations, thus a set of NLO terms of order $\mathcal{O}(\alpha_s^2)$ were abandoned; (ii) the LO BK approximation was used in the replacement of the first derivative in the expansions of the ``real" $S$-matrix\footnote{The ``real" means a really measuring the scattering of the soft gluon.} when the ciBK equation in $Y$-representation was transformed to $\eta$-representation\cite{DIMST19}, which leads to lose precision at the level of NLO accuracy.

In this paper, we shall derive the next-to-leading order BK equation in $\eta$-representation by including the running coupling corrections during the expansions of the ``real" $S$-matrix. To see the significance of the corrections of the running coupling, we firstly recall the derivation of the ciBK-$\eta$ equation with the LO BK approximation in the replacement of the first derivative in the expansion of the ``real" $S$-matrix. Second, we derive an extended collinearly-improved Balitsky-Kovchegov equation in $\eta$-representation by emphasizing the running coupling corrections in the expansion of the ``real" $S$-matrix. We obtain an extended ``canonical'' Balitsky-Kovchegov equation (exBK-$\eta$) whose evolution kernel is modified by the running coupling corrections as compared to the caBK-$\eta$ equation. The exBK-$\eta$ equation is analytically solved in saturation region. Its analytic solution shows that the exponent of the $S$-matrix has a linear dependence on rapidity instead of a quadratic rapidity dependence in the caBK-$\eta$ case, which obeys a similar law as results in $Y$-representation that the evolution speed of the dipole amplitude is suppressed by the running coupling corrections. We compare our extended collinearly-improved Balitsky-Kovchegov equation in $\eta$ with its original form to see how big difference is. It is easy to find that there are eight extra terms resulting from the running coupling corrections, which indicate that the precision of the expansion of the $S$-matrix has a significant impact on the evolution equation.

Finally, we numerically solve the evolution equations in $\eta$-representation to test the analytic outcomes mentioned above. The numerical results confirm our analytic findings. The saturation exponents $\bar{\lambda}$ are extracted from the solutions of the LO BK, caBK-$\eta$ and exBK-$\eta$ equations. As expected, the running coupling effect has a large influence on the evolution speed of the front, which largely suppresses the rapidity evolution of the dipole amplitude.

\section{Leading order, and running coupling BK equations in $Y$-representation}
\label{sec:loandrc}

In order to collect the basic elements of the dipole evolution equations, we give a brief recall of the LO BK and rcBK equations in $Y$-representation. These two equations shall be used in the Taylor expansion of the ``real" $S$-matrix in the derivation of the collinearly-improved BK equation in $\eta$-representation in the next section.

\subsection{Leading order BK equation}
\label{sec:lobk}

We consider the high energy scattering between a dipole which is consisted of a quark-antiquark pair moving towards the positive direction of the longitudinal axis with momentum ($p^{+}$, $p^{-}$, $\bm{p}$), and a hadronic target moving along the negative direction with momentum ($p_0^{+}$, $p_0^{-}$, $\bm{p}_0$). The scattering is treated in the eikonal approximation, thus the transverse coordinates of the quark ($\bm{x}$) and the antiquark ($\bm{y}$) are not modified by the collision. One can write the dipole scattering matrix as a correlator of two Wilson lines\cite{BC08}
\be
S_{\bx\by}(Y)= \frac{1}{N_c}\big\langle \mathrm{Tr}\{U(\bm{x})U^{\dagger}(\bm{y})\}\big\rangle_Y,
\label{S_matrix}
\ee
where the $\langle\cdots\rangle_Y$ means the average over target gluon field configurations at $Y$. Here, the $Y$ is the rapidity difference between the dipole and hadronic target
\be
Y=\ln\frac{p^{+}}{p_0^{+}} = \ln\frac{2p^{+}p_0^{-}}{Q_0^2} = \ln\frac{s}{Q_0^2},
\ee
with the center of mass energy squared $s=2p^{+}p_0^{-}$ and the typical momentum of the target $Q_0$.
The $U$ in Eq.(\ref{S_matrix}) is the time ordered Wilson line
\be
U(\bm{x})=\mathrm{P}\exp\Big[ig\int \dif x^{-}A^{+}(x^{-},\bm{x})\Big],
\ee
with $A^{+}(x^{-},\bm{x})$ as the gluon field of the hadronic target.

In the mean field approximation, the rapidity evolution of the $S$-matrix satisfies the BK equation\cite{B,K}
\be
\frac{\partial}{\partial Y} S_{\bx\by}(Y) = \frac{\bar{\alpha}_s}{2\pi} \int \dif^2\bm{z}\frac{(\bm{x}-\bm{y})^2}{(\bm{x}-\bm{z})^2(\bm{z}-\bm{y})^2}
\left [S_{\bx\bz}(Y)S_{\bz\by}(Y) - S_{\bx\by}(Y) \right],
\label{LOBK}
\ee
with $\bar{\alpha}_s = \alpha_sN_c/\pi$, and $\bm{x},\bm{y},\bm{z}$ as the transverse coordinates of the quark, antiquark, and emitted gluon, respectively. In large $N_c$ limit, the Eq.(\ref{LOBK}) depicts the parent dipole ($\bm{x}$, $\bm{y}$) evolved into two daughter dipoles ($\bm{x}$, $\bm{z}$) and ($\bm{z}$, $\bm{y}$) as the rapidity increasing. The first term in the right hand side of Eq.(\ref{LOBK}) is called as ``real'' term which describes the two daughter dipoles scattering with the target simultaneously, thus it is a non-linear term. The second term in the right hand side of Eq.(\ref{LOBK}) is referred as ``virtual'' term which depicts the survival probability of the original dipole at the time of scattering. Note that the BK equation resums only the leading logarithmic $\alpha_s\ln(1/x)$ corrections in the fixed coupling case, thus it is a LO evolution equation.

In the later section of this paper, we shall be interested in the limiting form of the $S$-matrix in saturation region where the dipole has very large size, such that $r^2Q_s^2\gg1$. Here, the $Q_s$ is the typical transverse momentum of the saturated gluons and is rapidly increasing with $Y$. For later comparison, we analytically solve the Eq.(\ref{LOBK}) in the following. In saturation regime, the $S$-matrix approaches the black-disk limit, $S\rightarrow0$. Therefore, the quadratic term in $S$ in Eq.(\ref{LOBK}) can be neglected. Moreover, the saturation condition implies that the two daughter dipoles are larger than the typical transverse size $r_s\sim1/Q_s$. Therefore, the Eq.(\ref{LOBK}) can reduce to
\be
\frac{\partial S(r, Y)}{\partial Y} \simeq
           -\frac{\bar{\alpha}_s}{2\pi}\int_{1/Q_s}^{r} \frac{\dif^2r_1 r^2}{r_1^2r_2^2}
           S(r, Y).
\label{simLOBK}
\ee
with $r=|\bm{x}-\bm{y}|$, $r_1=|\bm{x}-\bm{z}|$, $r_2=|\bm{z}-\bm{y}|$ be the transverse size of the parent dipole, and two daughter dipoles, respectively. Further, one can find that the integral in Eq.(\ref{simLOBK}) is governed by the case if one of the daughter dipoles is much smaller than the parent dipole, $r_1\ll r$ and $r_2\sim r$, or $r_2\ll r$ and $r_1\sim r$\cite{Mueller,Xiang09}. We select to work in the case $r_2\sim r$, the Eq.(\ref{simLOBK}) can be rewritten as
\be
\frac{\partial S(r, Y)}{\partial Y} \simeq
           -\bar{\alpha}_s\int_{1/Q_s^2}^{r^2} \frac{\dif r_1^2}{r_1^2}
           S(r, Y),
\label{eq_LOBK_apf}
\ee
where the right hand side includes a factor of $2$ to take into account that the smaller dipole with size $r_1$ can be any one of the two daughter dipoles. Now we carry out the integrals in Eq.(\ref{eq_LOBK_apf}), which yield the analytic solution of the LO BK equation as
\cite{Mueller,Xiang09,Levin-Tuchin,Xiang17},
\be
S(r, Y)=\exp\left[-\frac{\lambda\abar^2}{2}(Y-Y_0)^2\right]S(r, Y_0),
\label{SolLOBK}
\ee
where we have used $\ln[r^2Q_s^2(Y)]\simeq \lambda\abar(Y-Y_0)$ with $Y_0$ the rapidity scale at which $Q_s^2(Y_0)=1/r^2$. The Eq.(\ref{SolLOBK}) is usually called as Levin-Tuchin formula, since it was firstly derived by them in Ref.\cite{Levin-Tuchin}. From the above derivation, we know that the exponent in Eq.(\ref{SolLOBK}) is known only to leading double logarithmic accuracy, which means that the sub-leading terms are not under control. In addition, the exponent of the $S$-matrix has a quadratic dependence on rapidity, which renders the $S$-matrix too small at large rapidities, in the other words, the rapidity evolution speed of the dipole amplitude $N$ is too fast, with $N=1-S$. It has been recognized that the aforemention drawbacks are the reasons why the LO BK equation is insufficient to describe the experimental data at HERA\cite{JS,JM,Levin16}. Therefore, one has to include the NLO corrections to the LO BK evolution equation, such as the running coupling effect which can suppress the evolution speed of the dipole amplitude by modifying the evolution kernel of the BK equation\cite{Bnlo,KW}.


\subsection{Running coupling BK equation}
\label{sec:rcBK}
It is known that the LO BK equation discussed above considers only the resummation of leading logarithmic $\alpha_s\ln(1/x)$ corrections with a fixed coupling. Beyond the leading logarithmic approximation, there was a significant progress in the BK evolution equation via the resummation of $\alpha_sN_f$ to all orders, which is called as the running coupling corrections. The running coupling Balitsky-Kovchegov (rcBK) equation have been derived independently by Balitsky in Ref.\cite{Bnlo} and Kovchegov and Weigert in Ref.\cite{KW}. This two groups obtained an analogous structure of the rcBK equation but with different evolution kernels. In this study, we shall use the Balitsky version of kernel, since it is favored by the HERA data\cite{JS}. Note that we don't plan to give a details of the derivation of the rcBK equation, it is out of interesting of this paper.
The rcBK equation reads
\be
\frac{\partial S(r, Y)}{\partial Y} = \int\,\dif^2 r_1
  \,K^{\mathrm{rc}}(r, r_1, r_2)
  \left[S(r_1, Y)\,S(r_2, Y)-S(r, Y)\right],
\label{rcBK}
\ee
where $K^{\mathrm{rc}}(r, r_1, r_2)$ is the running coupling evolution kernel\cite{Bnlo}
\be
K^{\mathrm{rc}}(r, r_1, r_2) = \frac{N_c\alpha_s(r_{\mathrm{min}}^2)}{2\pi^2}\left[\frac{r^2}{r_1^2r_2^2} + \frac{1}{r_1^2}\left(\frac{\alpha_s(r_1^2)}{\alpha_s(r_2^2)}-1\right)
+ \frac{1}{r_2^2}\left(\frac{\alpha_s(r_2^2)}{\alpha_s(r_1^2)}-1\right)\right],
\label{Bal_rc}
\ee
with $r_{\mathrm{min}}=\mathrm{min}\{r,r_1,r_2\}$. There are several running coupling prescriptions in the literature\cite{Bnlo,IMMST2,Xiang20,Albacete17,Cepila19}. At the very beginning, the argument of the running coupling $\alpha_s$ in the rcBK equation was interpreted to the transverse size of the parent dipole. Recently, it was found that the size of the smallest dipole is a proper argument of the coupling, and the smallest running coupling prescription is favored by the HERA data than others\cite{IMMST2}. So, we shall use the smallest dipole running coupling prescription in this study. In addition, the running coupling at one loop accuracy is used
\be
\alpha_s(r^2) = \frac{1}{b\ln\big(\frac{1}{r^2\Lambda^2}\big)},
\label{oneloop}
\ee
with $b=(11N_c-2N_f)/12\pi$.

Now, let us turn to analytically solve the rcBK equation. As it was done in the previous section, we solve the rcBK equation in saturation region where the dipoles have very large size, $r$, $r_1$, $r_2\geq1/Q_s$. The integral over the $r_1$ in Eq.(\ref{rcBK}) is governed by the case if one of the daughter dipoles is much smaller than the parent one, while the rest of dipole has a similar size as the parent dipole, that is $r_1\ll r$ and $r_2\sim r$, or $r_2\ll r$ and $r_1\sim r$. We choose to work in the first case, the Eq.(\ref{rcBK}) reduces to
\be
\frac{\partial S(r, Y)}{\partial Y} = \frac{1}{\pi} \int_{1/Q_s}^r\,
   \frac{\dif^2 r_1\abar(r_1^2)}{r_1^2}
  \left[S(r_1, Y)\,S(r_2, Y)-S(r, Y)\right],
\label{rcBKf}
\ee
where the right hand size includes a factor of $2$ due to the fact that the smaller dipole can come from either of the two daughter dipoles. In saturation region, the $S$-matrix approaches the black-disk limit, one has $S\rightarrow0$, thus the quadratic term in $S$ in Eq.(\ref{rcBKf}) can be discarded.
The Eq.(\ref{rcBKf}) simplifies to
\be
\frac{\partial S(r,Y)}{\partial Y}\simeq -\int_{1/Q_s^2}^{r^2}\,
  \,\frac{\dif r_1^2\bar{\alpha}_s(r_1^2)}{r_1^2}S(r, Y).
\label{SKW_sr}
\ee
Substituting Eq.(\ref{oneloop}) into Eq.(\ref{SKW_sr}), and performing the integrals over $r_1$ and $Y$, we can get the analytic solution of the rcBK equation as\cite{Xiang09,Xiang17}
\be
S(r,Y) = \exp\left\{-\frac{N_c}{b\pi}(Y-Y_0)\left[\ln\left(\frac{\sqrt{\lambda'(Y-Y_0)}}{\ln\frac{1}{r^2\Lambda^2}}\right)-\frac{1}{2}\right]\right\}S(r, Y_0),
\label{solrcBK}
\ee
where the NLO saturation momentum is used
\be
\ln\frac{Q_s^2}{\Lambda^2} = \sqrt{\lambda'(Y-Y_0)} + \mathcal{O}(Y^{1/6}).
\label{SM}
\ee
If one compares the solution of the rcBK equation, Eq.(\ref{solrcBK}), with the solution of the LO BK equation, Eq.(\ref{SolLOBK}). It is easy to find that the quadratic rapidity dependence in the exponent of the $S$-matrix is replaced by the linear rapidity dependence due to the running coupling corrections. This outcome indicates that the evolution speed of the dipole amplitude is slowed down by the running coupling corrections. This finding is consistent with the theoretical expectations\cite{Bnlo,AK07}. Furthermore, the phenomenological applications of the rcBK equation at HERA energies show that the rcBK equation gives a more reasonable description of the experimental data than the LO BK equation\cite{JS,JM}.

\section{Collinearly-improved BK equation in $\eta$-representation}
\label{sec:loandrc}
In previous section, all the dipole evolution equations are studied in the $Y$-representation. In this section, we shall discuss the dipole evolution equations in $\eta$-representation due to two key reasons. On one hand, a recent study in Ref.\cite{DIST20} realized that the rapidity $\eta=\ln(1/x)$ of the hadronic target is the physical rapidity used in the DIS experiments at HERA rather than the projectile rapidity $Y$. On the other hand, it was found that the reason for the instability of the full NLO BK equation is a consequence of the wrong choice of the ``evolution time'', this refers to the choice of the rapidity variable\cite{DIMST19}. However, if one simply transforms the ciBK-$Y$ equation to $\eta$-representation by change of variable $\eta = Y - \rho$, the instability problem is still existence due to NLO corrections enhanced by double collinear logarithms, but not as severe as for the corresponding issue in $Y$. Based on the previous experience with the full NLO BK equation in $Y$, where a similar problem was identified and eventually cured by enforcing the time-ordering constrains on the successive gluon emissions, one can know that the ordering of the successive emissions in longitudinal momentum should be enforced in order to solve instability problem in $\eta$-representation. By doing this, the ciBK-$\eta$ was obtained in Ref.\cite{DIMST19}, which can directly apply to phenomenology and supposes to give a better description of the HERA data than the other evolution equations in the literature. However, a very recent study in Ref.\cite{Beuf20} showed that three different evolution equations (kinematical constraint BK (kcBK)\cite{Beuf14}, ciBK\cite{IMMST2}, and caBK-$\eta$\cite{DIMST19} equations) result in a very similar description of the HERA data. Note that the first two equations are presented in $Y$-representation, while the last equation is given in $\eta$-representation. It is easy to understand that the kcBK and ciBK equations give a equally good depiction of the data, since they are equivalent in the sense that the resummation of the leading double transverse logarithms is concerned. The reason why the caBK-$\eta$ does not give a superior description of the data, could possibly be attributed to the insufficient accuracy used in the expansion of the ``real" $S$-matrix when the caBK-$\eta$ equation was derived. In this section, we shall derive the collinearly-improved BK equation in $\eta$ at the level of running coupling when the ``real" $S$-matrix is expanded. An extended collinearly-improved BK equation in $\eta$ is obtained, which has the same structure as the ciBK-$\eta$ equation but with a running coupling modified kernel. In the next section, we use the numerical method to solve the caBK-$\eta$ and exBK-$\eta$ equations. The numerical results show that the running coupling corrections play a significant role in the suppression of the evolution of the dipole amplitude.

\subsection{Collinearly-improved BK equation in $\eta$: LO BK approximation in expansion of $S$-matrix}
To obtain the collinearly-improved BK equation in $\eta$, we follow the same strategy used in Ref.\cite{DIMST19} where the change of variable is employed to transform the ciBK equation from the $Y$-representation to the $\eta$-representation.

We start with the full NLO BK equation in $Y$-representation\cite{Bnlo}
\begin{align}
 \hspace*{0.5cm}
 \frac{\partial S_{\bx\by}(Y)}{\partial Y} &= \,
 \frac{\abar}{2 \pi}
\int\dif^2 \bz\,
 \cdot(K_0 + K_q + K_g)\cdot
 \Big(S_{\bx\bz}(Y) S_{\bz\by}(Y) - S_{\bx\by}(Y) \Big)
 \nn
 & \hspace*{0.5cm} +\frac{\abar^2}{8\pi^2}
 \int\dif^2 \bu \,\dif^2 \bz\cdot K_1 \cdot
\Big(S_{\bx\bu}(Y) S_{\bu\bz}(Y) S_{\bz\by}(Y) - S_{\bx \bu}(Y) S_{\bu \by}(Y)\Big)
\nn
 & \hspace*{0.5cm} + \frac{\abar^2}{8\pi^2}\frac{N_f}{N_c}\int\dif^2 \bu \,\dif^2 \bz
\cdot K_f\cdot\Big(S_{\bx\bz}(Y)S_{\bu\by}(Y)-S_{\bx\bu}(Y)S_{\bu\by}(Y)\Big)
\label{fnlobkY}
\end{align}
with
\begin{align}
K_0 = \frac{(\bx-\by)^2}{(\bx-\bz)^2(\bz-\by)^2},
\label{K0}
\end{align}
\begin{align}
K_{q} =  \frac{(\bx-\by)^2}{(\bx-\bz)^2(\bz-\by)^2}\abar
 \lt[b \ln (\bx \minus \by)^2 \mu^2
 - b\frac{(\bx \minus\bz)^2 - (\by \minus\bz)^2}{(\bx \minus \by)^2}
 \ln\frac{(\bx \minus\bz)^2}{(\by \minus\bz)^2}\rt],
 \label{Kq}
\end{align}
\begin{align}
K_{g} &=  \frac{(\bx-\by)^2}{(\bx-\bz)^2(\bz-\by)^2}\abar\bigg[
\frac{67}{36} - \frac{\pi^2}{12} -\frac{5N_f}{18N_c}
-\frac{1}{2}\ln \frac{(\bx \minus\bz)^2}{(\bx \minus\by)^2} \ln \frac{(\by \minus\bz)^2}{(\bx \minus\by)^2}\bigg],
\label{Kg}
\end{align}
\begin{align}
K_1 &= \frac{1}{(\bu-\bz)^4}\bigg\{\minus 2
 + \frac{(\bx \minus\bu)^2 (\by \minus\bz)^2 +
 (\bx \minus \bz)^2 (\by \minus \bu)^2
 - 4 (\bx \minus \by)^2 (\bu \minus \bz)^2}{(\bx \minus \bu)^2 (\by  \minus \bz)^2 - (\bx \minus \bz)^2 (\by \minus \bu)^2}
\ln \frac{(\bx \minus \bu)^2 (\by  \minus \bz)^2}{(\bx \minus \bz)^2 (\by \minus \bu)^2}\nn
 &\hspace*{2.1cm}+ \frac{(\bx \minus \by)^2 (\bu \minus \bz)^2}{(\bx \minus \bu)^2 (\by  \minus \bz)^2}
 \left[1 + \frac{(\bx \minus \by)^2 (\bu \minus \bz)^2}{(\bx \minus \bu)^2 (\by  \minus \bz)^2 - (\bx \minus \bz)^2 (\by \minus \bu)^2} \right]
\ln \frac{(\bx \minus \bu)^2 (\by  \minus \bz)^2}{(\bx \minus \bz)^2 (\by \minus \bu)^2}\bigg\},
\label{K1}
\end{align}
and
\begin{align}
K_f = \frac{1}{(\bu-\bz)^4} \bigg[2-\frac{(\bx \minus\bu)^2 (\by \minus\bz)^2 +
 (\bx \minus \bz)^2 (\by \minus \bu)^2
 - (\bx \minus \by)^2 (\bu \minus \bz)^2}{(\bx \minus \bu)^2 (\by  \minus \bz)^2 - (\bx \minus \bz)^2 (\by \minus \bu)^2}
\ln \frac{(\bx \minus \bu)^2 (\by  \minus \bz)^2}{(\bx \minus \bz)^2 (\by \minus \bu)^2}
 \bigg].
\label{Kf}
\end{align}
From Eq.(\ref{fnlobkY}), one can see that the full NLO BK equation has two main changes in the structure as compared to the LO BK equation. The first term in the right hand side receives corrections from quark loops ($K_q$) and gluon loop ($K_g$). The last two terms in the right hand side refer to partonic fluctuations involving two additional partons except the original parent partons (quark and antiquark). In large $N_c$ limit, they are corresponding to two consecutive emissions, the original parent dipole ($\bx, \by$) emits a gluon at transverse coordinate $\bu$, which is equivalent to two daughter dipoles ($\bx, \bu$) and ($\bu,\by$). Then the dipole ($\bu,\by$) emits a gluon at transverse coordinate $\bz$, which yields the dipoles ($\bu, \bz$) and ($\bz,\by$).

\subsubsection{The ``canonical" BK equation}
\label{sec:caBK}
In order to transform the Eq.(\ref{fnlobkY}) from $Y$-representation into $\eta$-representation, we need to change variables in terms of
\be
Y = \eta + \rho.
\label{eta2Y}
\ee
Now, we can rewrite the $S$-matrices in $\eta$ representation as
\begin{align}
S_{\bx\by}(Y) = S_{\bx\by}(\eta+\rho)
\equiv \bar{S}_{\bx\by}(\eta),
\label{Sxyeta}
\end{align}
\begin{align}
S_{\bx\bz}(Y) = S_{\bx\bz}(\eta+\rho) =
S_{\bx\bz}\left(\eta +\ln\frac{(\bx-\bz)^2}{(\bx-\by)^2}+ \rho_{\bx\bz}\right) =
\bar{S}_{\bx\bz} \left(\eta +\ln\frac{(\bx-\bz)^2}{(\bx-\by)^2}\right),
\label{Sxzeta}
\end{align}
\begin{align}
S_{\bz\by}(Y) = S_{\bz\by}(\eta+\rho) =
S_{\bz\by}\left(\eta +\ln\frac{(\by-\bz)^2}{(\bx-\by)^2}+ \rho_{\bz\by}\right) =
\bar{S}_{\bz\by} \left(\eta +\ln\frac{(\by-\bz)^2}{(\bx-\by)^2}\right),
\label{Szyeta}
\end{align}
where we have used
\begin{align}
\rho =\ln\Big(\frac{Q^2}{Q_0^2}\Big)=\ln\bigg(\frac{1}{(\bx-\by)^2 Q_0^2}\bigg)=\ln\bigg(\frac{(\bx-\bz)^2}{(\bx-\bz)^2(\bx-\by)^2 Q_0^2}\bigg)=\ln\bigg(\frac{(\bx-\bz)^2}{(\bx-\by)^2}\bigg)+\rho_{\bx\bz},
\label{rho0}
\end{align}
and
\begin{align}
\rho =\ln\Big(\frac{Q^2}{Q_0^2}\Big)=\ln\bigg(\frac{1}{(\bx-\by)^2 Q_0^2}\bigg)=\ln\bigg(\frac{(\bz-\by)^2}{(\bz-\by)^2(\bx-\by)^2 Q_0^2}\bigg)=\ln\bigg(\frac{(\bz-\by)^2}{(\bx-\by)^2}\bigg)+\rho_{\bz\by}.
\label{rho0}
\end{align}
When one works at NLO in $\abar$, one can expand out the ``real" $S$-matrices ($S_{\bx\bz}$ and $S_{\bz\by}$) in Taylor series, since the rapidity shift in the argument of $S$-matrices is typically much smaller than $\eta$ itself. The expansions of the $S_{\bx\bz}$ and $S_{\bz\by}$ can be expressed as
\begin{align}
S_{\bx\bz}(Y)=\bar{S}_{\bx\bz} \left(\eta +\ln\frac{(\bx-\bz)^2}{(\bx-\by)^2}\right)
&\simeq
\bar{S}_{\bx\bz}(\eta)
+ \ln\frac{(\bx-\bz)^2}{(\bx-\by)^2}
\frac{\partial \bar{S}_{\bx\bz}(\eta)}{\partial \eta},
\label{expandxz}
\end{align}
and
\begin{align}
S_{\bz\by}(Y)=\bar{S}_{\bz\by} \left(\eta +\ln\frac{(\by-\bz)^2}{(\bx-\by)^2}\right)
&\simeq
\bar{S}_{\bz\by}(\eta)
+ \ln\frac{(\by-\bz)^2}{(\bx-\by)^2}
\frac{\partial \bar{S}_{\bz\by}(\eta)}{\partial \eta}.
\label{expandzy}
\end{align}
To the order of interesting, we need only to keep the first non-trivial term in the above expansions, since each $\partial S/\partial\eta$ is formally suppressed by a power of $\abar$. For the derivative terms in Eqs.(\ref{expandxz}) and (\ref{expandzy}), in this subsection we use the LO BK equation to evaluate them as what was done in Ref.\cite{DIMST19},
\be
S_{\bx\bz}(Y)=\bar{S}_{\bx\bz} \left(\eta +\ln\frac{(\bx-\bz)^2}{(\bx-\by)^2}\right)
\simeq\, \bar{S}_{\bx\bz}(\eta)+
\frac{\abar}{2\pi}\int\frac{ \dif^2 \bu\,(\bx\minus\bz)^2}{(\bx \minus\bu)^2 (\bu \minus \bz)^2}\,
\ln\frac{(\bx-\bz)^2}{(\bx-\by)^2}
\left[\bar{S}_{\bx\bu}(\eta) \bar{S}_{\bu\bz}(\eta) - \bar{S}_{\bx\bz}(\eta) \right],
\label{expandxz_LO}
\ee
and
\be
S_{\bz\by}(Y)=\bar{S}_{\bz\by} \left(\eta +\ln\frac{(\by-\bz)^2}{(\bx-\by)^2}\right)
\simeq\, \bar{S}_{\bz\by}(\eta)+
\frac{\abar}{2\pi}\int\frac{ \dif^2 \bu\,(\by\minus\bz)^2}{(\by \minus\bu)^2 (\bu \minus \bz)^2}\,
\ln\frac{(\by-\bz)^2}{(\bx-\by)^2}
\left[\bar{S}_{\bz\bu}(\eta) \bar{S}_{\bu\by}(\eta) - \bar{S}_{\bz\by}(\eta) \right].
\label{expandzy_LO}
\ee
From the above equations, one can see that the rapidity shift in the argument of $S$-matrices is equivalent to adding a term of order $\mathcal{O}(\abar)$.
We would like to point out that the LO BK equation is a rough approximation to estimate the derivative terms in Eqs.(\ref{expandxz}) and (\ref{expandzy}).
Actually, the LO BK equation is insufficient due to its lower precision. To reach the interested order of accuracy, one should use at least the level of rcBK equation to evaluate the derivative terms, which shall study in the next subsection.

Now, substituting Eqs.(\ref{expandxz_LO}) and (\ref{expandzy_LO}) into Eq.(\ref{fnlobkY}), one can get a semi-finished collinearly-improved BK equation\footnote{The reason why we call it as semi-finished equation is that it is still a unstable equation, which needs to do the resummations of the double collinear logarithms to totally get rid of the instability.} in $\eta$\cite{DIMST19}
\begin{align}
 \hspace*{0.5cm}
 \frac{\partial S_{\bx\by}(\eta)}{\partial \eta} &= \,
\frac{\abar}{2 \pi}
 \int
 \dif^2\cdot K_0\cdot\,
\Big(\bar{S}_{\bx\bz}(\eta)\bar{S}_{\bz\by}(\eta)- \bar{S}_{\bx\by}(\eta)\Big)
\nn
& \hspace*{0.5cm}
+\frac{\abar}{2 \pi}\int
\dif^2\bz\cdot (K_q + K_g)\cdot \,
 \Big(\bar{S}_{\bx\bz}(\eta) \bar{S}_{\bz\by}(\eta) - \bar{S}_{\bx\by}(\eta) \Big)
 \nn
& \hspace*{0.5cm} + \frac{\abar^2}{2\pi^2}
 \int \dif^2 \bz\, \dif^2 \bu\,\cdot K_l\cdot
  \bar{S}_{\bx\bu}(\eta)
 \Big(\bar{S}_{\bu\bz}(\eta) \bar{S}_{\bz\by}(\eta) - \bar{S}_{\bu\by}(\eta) \Big)
 \nn
 & \hspace*{0.5cm} +\frac{\abar^2}{8\pi^2}
 \int\dif^2 \bu \,\dif^2 \bz\cdot K_1 \cdot
\Big(\bar{S}_{\bx\bu}(\eta) \bar{S}_{\bu\bz}(\eta) \bar{S}_{\bz\by}(\eta) - \bar{S}_{\bx \bu}(\eta) \bar{S}_{\bu \by}(\eta)\Big)
\nn
 & \hspace*{0.5cm} + \frac{\abar^2}{8\pi^2}\frac{N_f}{N_c}\int\dif^2 \bu \,\dif^2 \bz
\cdot K_f\cdot\Big(\bar{S}_{\bx\bz}(\eta)\bar{S}_{\bu\by}(\eta)-\bar{S}_{\bx\bu}(\eta)\bar{S}_{\bu\by}(\eta)\Big),
\label{fnlobk1}
\end{align}
with \be
K_l = \frac{(\bx \minus \by)^2}{(\bx \minus \bu)^2 (\bu \minus \bz)^2 (\bz \minus \by)^2}\ln\frac{(\bu-\by)^2}{(\bx-\by)^2}.
\ee
The third term in the right hand side of Eq.(\ref{fnlobk1}) is resulting from the expansion of the ``real" $S$-matrices in terms of rapidity shift with the LO BK approximation in the evaluating of the derivative term.
Note that the rapidity shift is neglected in the all NLO terms when Eq.(\ref{fnlobk1}) is derived. All the terms of order $\mathcal{O}(\abar^2)$ in Eq.(\ref{fnlobkY}) are simply replaced like $S_{\bx\bz}(Y)\rightarrow\bar{S}_{\bx\bz}(\eta)$, since the rapidity shift take a contribution of order $\mathcal{O}(\abar)$ which renders all the NLO terms of order $\mathcal{O}(\abar^3)$. While we are only interested in the terms up to the order of $\mathcal{O}(\abar^2)$, thus all the terms beyond $\abar^2$ are abandoned in this paper. Moreover, two mathematical tricks are used when Eq.(\ref{fnlobk1}) is derived, (i) the property that the LO term is invariant under $\bx - \bz\rightarrow \bz-\by$, is exploited to combine some terms; (ii) the integral variables in the third term in the right hand side of Eq.(\ref{fnlobk1}) is relabelled in terms of $\bu\leftrightarrow\bz$ in order to keep consistence with the physics picture mentioned above.

The Eq.(\ref{fnlobk1}) is a NLO evolution equation in $\eta$-representation, which is a local equation in rapidity. By comparing Eq.(\ref{fnlobk1}) with Eq.(\ref{fnlobkY}), one can see that the difference between them is only by an extra term (resulting from the change of variable) in the third line in the right hand side of Eq.(\ref{fnlobk1}). Originally, one anticipates that the change of variable from $Y$ to $\eta$ can eliminate the instabilities occurred in Eq.(\ref{fnlobkY}). As expected that the instabilities caused by the violations of time-ordering (double anti-collinear logarithms) are disappeared in Eq.(\ref{fnlobk1}), since the time-ordering property is automatically guaranteed in the $\eta$ evolution. Unfortunately, it has been shown that the change of variable triggers off another type of instabilities associated with double collinear logarithms\cite{DIMST19}.

To cure the instabilities mentioned above, it is known that the successive gluon emissions during the rapidity evolution have to be simultaneously ordered in lifetime and longitudinal momentum,
\be
\tau_p \gg \tau_k \gg \tau_0,
\label{ltcst}
\ee
and
\be
 p^{+}\gg k^{+} \gg p_0^{+} ~~~\Rightarrow~~~ \frac{2p^{+}}{Q^2}Q^2 \gg \frac{2k^{+}}{\bm{k}^2}\bm{k}^2 \gg \frac{2p_0^{+}}{Q_0^2}Q_0^2 ~~~\Rightarrow~~~ \tau_pQ^2 \gg \tau_{\bm{k}}\bm{k}^2 \gg \tau_0Q_0^2,
\label{lmcst}
\ee
where the ($p^{+}$, $p^{-}$, $\bm{p}$), ($p_0^{+}$, $p_0^{-}$, $\bm{p}_0$), and ($k^{+}$, $k^{-}$, $\bm{k}$) denote the light cone momenta of the projectile, target and emitted gluon, respectively. The first constraint is automatically satisfied as mentioned above. However, the second constraint may be violated when the radiated gluon is either too soft ($\bm{k}^2\ll Q_0^2$) or too hard ($\bm{k}^2\gg Q^2$). So, one has to put the constraint on the evolution equation. Before doing that, we need to rewrite the constraint, Eq.(\ref{lmcst}), in a proper form. As we know
\be
\rho = \ln\Big(\frac{Q^2}{Q_0^2}\Big), ~~~~~~ Y = \ln\Big(\frac{p^{+}}{p_0^{+}}\Big),
\label{rho}
\ee
and
\be
\rho_1 = \ln\Big(\frac{\bm{k}^2}{Q_0^2}\Big), ~~~~~~ Y_1 = \ln\Big(\frac{k^{+}}{p_0^{+}}\Big).
\label{rho1}
\ee
Thus, the target rapidities can be expressed as
\be
\eta = Y - \rho
     = \ln\Big(\frac{p^{+}}{p_0^{+}}\Big) - \ln\Big(\frac{Q^2}{Q_0^2}\Big)
     = \ln\Big(\frac{\tau_p}{\tau_0}\Big),
\label{eta}
\ee
and
\be
\eta_1 = Y_1 -\rho_1
       = \ln\Big(\frac{k^{+}}{p_0^{+}}\Big) - \ln\Big(\frac{\bm{k}^2}{Q_0^2}\Big)
       = \ln\Big(\frac{\tau_k}{\tau_0}\Big).
\label{eta_1}
\ee
Using Eqs.(\ref{eta}) and (\ref{eta_1}), one can rewrite the constraint, Eq.(\ref{lmcst}), as
\begin{align}
\eta - \ln (\frac{\bk^2}{Q^2})\gg \eta_1 \gg \ln (\frac{Q^2_0}{\bk^2} ).
\label{etacst1}
\end{align}
Moreover, according to the lifetime constraint in Eq.(\ref{ltcst}), one can get
\be
\eta \gg \eta_1 \gg 0.
\label{etacst2}
\ee
Combining the two constraints, Eqs.(\ref{etacst1}) and (\ref{etacst2}), we obtain the final rapidity constraint as
\begin{align}
{\rm min}\left\{\eta, \eta-\ln\frac{\bk^2}{Q^2}\right\} \, >\,\eta_1\,>\,
{\rm max}\left\{ 0, \ln\frac{Q_0^2}{\bk^2}\right\}.
\end{align}
By using Eq.(\ref{rho}) and (\ref{rho1}), the above equation can be written in another form as
\be
\Theta(-\rho_1)|\rho_1|\ll \eta_1 \ll\eta-\Theta(\rho_1-\rho)(\rho_1-\rho),
\label{fcst}
\ee
which is a proper form, and can be directly used.

Now, we apply the constraint, Eq.(\ref{fcst}), to the integral form of BK equation and get
\begin{align}
\bar S_{\bx\by}(\eta)= S^{(0)}_{\bx\by}+
\frac{\abar}{2\pi}
\int \frac{\dif^2 \bz \,(\bx\minus\by)^2}{(\bx \minus\bz)^2 (\bz \minus \by)^2}\int\limits_
{\Theta(- \rho_1) |\rho_1|}^{\eta-\Theta(\rho_1- \rho)(\rho_1- \rho)}\dif \eta_1
\big[ \bar S_{\bx\bz}(\eta_1)
\bar S_{\bz\by}(\eta_1) \minus  \bar S_{\bx\by}(\eta_1) \big]
\label{BKeta}
\end{align}
which turns to differential format as
\begin{align}
\frac{\del \bar{S}_{\bx\by}(\eta)}{\del \eta} =
\frac{\abar}{2\pi}
& \int \frac{\dif^2 \bz \,(\bx\minus\by)^2}{(\bx \minus\bz)^2 (\bz \minus \by)^2}\,
\Theta\big(\eta \minus \delta_{\bx\by\bz}\big)\,\Theta\big(\eta \minus \Theta(-\rho_1) |\rho_1|\big)
\nn
&\times \big[\bar{S}_{\bx\bz}(\eta \minus \delta_{\bx\by\bz})\bar{S}_{\bz\by}(\eta \minus \delta_{\bx\by\bz})
\minus \bar{S}_{\bx\by}(\eta \minus \delta_{\bx\by\bz}) \big],
\label{fBKeta}
\end{align}
with the rapidity shift $\delta_{\bx\by\bz}$ as
\begin{align}
\delta_{\bx\by\bz}
= {\rm max}\left\{ 0,\ln \frac{(\bx\!-\!\by)^2} {{\rm min}
\{(\bx\!-\!\bz)^2,(\bz\!-\!\by)^2\}}\right\}.
\label{delta_xyz}
\end{align}
It has been checked that the Eq.(\ref{fBKeta}) has little scheme dependence on the prescription of rapidity shift. In this study, we choose to work with the ``canonical" one\cite{DIMST19}
\begin{align}
\bar{S}_{\bx\bz}(\eta \minus \delta_{\bx\by\bz})\bar{S}_{\bz\by}(\eta \minus \delta_{\bx\by\bz})
	\minus \bar{S}_{\bx\by}(\eta \minus \delta_{\bx\by\bz})\,\longrightarrow\,
	\bar{S}_{\bx\bz}(\eta \minus \delta_{\bx\bz; r})\bar{S}_{\bz\by}(\eta \minus \delta_{\bz\by;r})
	\minus \bar{S}_{\bx\by}(\eta),
\label{rapsft}
\end{align}	
with
\begin{align}
\delta_{\bx\bz;r} = {\rm max}\left\{ 0,\ln \frac{r^2}{(\bx\!-\!\bz)^2}\right\},
\label{delta1}
\end{align}
and
\begin{align}
\delta_{\bz\by;r} = {\rm max}\left\{ 0,\ln \frac{r^2}{(\bz\!-\!\by)^2}\right\}.
\label{delta2}
\end{align}

Substituting Eq.(\ref{rapsft}) into Eq.(\ref{fBKeta}), one gets the caBK-$\eta$ equation as\cite{DIMST19}
\begin{align}
\frac{\del \bar{S}_{\bx\by}(\eta)}{\del \eta} =
\frac{\abar}{2\pi}
\int\dif^2 \bz \cdot K_0\cdot\,
\Theta\big(\eta \minus \delta_{\bx\by\bz}\big)
\big[\bar{S}_{\bx\bz}(\eta \minus \delta_{\bx\bz;r})\bar{S}_{\bz\by}(\eta \minus \delta_{\bz\by;r})
\minus \bar{S}_{\bx\by}(\eta) \big],
\label{caBK}
\end{align}
which is a non-local evolution equation in rapidity $\eta$.

\subsubsection{Full collinear improved BK equation at $\mathcal{O}(\abar^2)$}
The caBK-$\eta$ equation can be generalized to full NLO accuracy by adding all the NLO corrections at $\mathcal{O}(\abar^2)$ in Eq.(\ref{fnlobk1}). Before adding those terms, we need to subtract the the $\mathcal{O}(\abar^2)$ piece in Eq.(\ref{caBK}). The $\mathcal{O}(\abar^2)$ piece in Eq.(\ref{caBK}) can be identified by expanding the $\bar{S}_{\bx\bz}(\eta \minus \delta_{\bx\bz;r})\bar{S}_{\bz\by}(\eta \minus \delta_{\bz\by;r})$ term as what we have done in Eqs.(\ref{expandxz_LO}) and (\ref{expandzy_LO}). One finds the $\mathcal{O}(\abar^2)$ piece as
\be
-\frac{\abar^2}{2\pi^2}
 \int \frac{\dif^2 \bz\, \dif^2 \bu\, (\bx \minus \by)^2 }{(\bx \minus \bu)^2 (\bu \minus \bz)^2 (\bz \minus \by)^2}\,
\delta_{\bu\by;r}\, \bar{S}_{\bx\bu}(\eta)
 \big[\bar{S}_{\bu\bz}(\eta) \bar{S}_{\bz\by}(\eta) - \bar{S}_{\bu\by}(\eta) \big].
\label{subT}
\ee
Adding the $\mathcal{O}(\abar^2)$ pieces from Eq.(\ref{fnlobk1}) to Eq.(\ref{caBK}) and subtracting the above $\mathcal{O}(\abar^2)$ piece, one obtains the ciBK-$\eta$ equation as\cite{DIMST19}
\begin{align}
 \hspace*{0.5cm}
 \frac{\partial S_{\bx\by}(\eta)}{\partial \eta} &= \,
\frac{\abar}{2 \pi}
 \int
\dif^2\bz\cdot K_0\cdot\,
\Theta\big(\eta \minus \delta_{\bx\by\bz}\big)
\Big(\bar{S}_{\bx\bz}(\eta \minus \delta_{\bx\bz;r})\bar{S}_{\bz\by}(\eta \minus \delta_{\bz\by;r})
\minus \bar{S}_{\bx\by}(\eta) \Big)
\nn
& \hspace*{0.5cm}
+\frac{\abar}{2 \pi}\int
\dif^2 \bz \cdot (K_q + K_g)\,
 \Big(\bar{S}_{\bx\bz}(\eta) \bar{S}_{\bz\by}(\eta) - \bar{S}_{\bx\by}(\eta) \Big)
 \nn
& \hspace*{0.5cm}+\frac{\abar^2}{2\pi^2}
 \int\dif^2 \bz\, \dif^2 \bu\,
 \cdot K_2 \cdot \bar{S}_{\bx\bu}(\eta)
 \Big(\bar{S}_{\bu\bz}(\eta) \bar{S}_{\bz\by}(\eta) - \bar{S}_{\bu\by}(\eta) \Big)
 \nn
 & \hspace*{0.5cm} +\frac{\abar^2}{8\pi^2}
 \int\dif^2 \bu \,\dif^2 \bz\cdot K_1 \cdot
\Big(\bar{S}_{\bx\bu}(\eta) \bar{S}_{\bu\bz}(\eta) \bar{S}_{\bz\by}(\eta) - \bar{S}_{\bx \bu}(\eta) \bar{S}_{\bu \by}(\eta)\Big)
\nn
 & \hspace*{0.5cm} + \frac{\abar^2}{8\pi^2}\frac{N_f}{N_c}\int\dif^2 \bu \,\dif^2 \bz
\cdot K_f\cdot\Big(\bar{S}_{\bx\bz}(\eta)\bar{S}_{\bu\by}(\eta)-\bar{S}_{\bx\bu}(\eta)\bar{S}_{\bu\by}(\eta)\Big),
\label{fnlobk}
\end{align}
with
\be
K_2 = K_l + \delta_{\bu\by;r}.
\ee
One can see that Eq.(\ref{fnlobk}) does not include double collinear logarithms now. These double logarithms are included in the first line in the right hand side of Eq.(\ref{fnlobk}) through the rapidity shift. In addition, the double anti-collinear logarithm term in kernel $K_g$ in the second line in the right hand side of Eq.(\ref{fnlobk}) is canceled by a relative piece generated by the integral over $\bu$ in the third line. All the unstable factors are under control in Eq.(\ref{fnlobk}). So, it is a stable equation which can directly apply to the phenomenological studies. However, a very recent study showed that the caBK-$\eta$ equation does not give a superior description of the HERA data than the kcBK and ciBK equations as the theoretical expectations\cite{Beuf20}. The reason why the outcomes resulting from the caBK-$\eta$ equation are not as desired, possibly comes from the insufficient accuracy of the expansion of the ``real" $S$-matrices (Eqs.(\ref{expandxz_LO}) and (\ref{expandzy_LO})) and the integral LO BK equation Eq.(\ref{BKeta}).

\subsubsection{Analytic solution to the caBK-$\eta$ equation in saturation region}
\label{sec:solcaBK}

Let us turn to analytically solve the caBK-$\eta$ equation in saturation region. In this regime, one of the daughter is much smaller than the other one, while the larger daughter dipole has comparable size as the parent dipole. As we know that the non-locality is only important for the $S$-matrix which is associated with smaller dipole. Thus, the Eq.(\ref{caBK}) can simplify to
\be
\frac{\del\bar{S}(r,\eta)}{\del\eta} \simeq 2\frac{\abar}{2\pi}\bar{S}(r,\eta)\int_{1/Q_s}^{r}\frac{\dif^2z}{z^2}\Theta\Big(\eta-\ln\frac{r^2}{z^2}\Big) \left[\bar{S}\Big(z,\eta-\ln\frac{r^2}{z^2}\Big)-1\right],
\label{ScaBK}
\ee
where a factor 2 is taken into account due to the fact that the smaller dipole can come from either of the two daughter dipoles, $Q_s$ is the saturation momentum which is associated with $\bar{Q}_s$ as $r^2Q_s^2=(r^2\bar{Q}_s^2)^{1/(1+\bar{\lambda})}$\cite{DIMST19}. For simplicity, we denote $z$ as the size of the smaller dipole in Eq.(\ref{ScaBK}).

To solve Eq.(\ref{ScaBK}) in saturation region, we follow two rules to do the calculations, (i) the saturation condition requires the dipole size $z$ larger than the typical size $1/Q_s$ (lower integral bound in Eq.(\ref{ScaBK})), which leads to the ``real" $S$-matrix $S(z,\eta-\ln\frac{r^2}{z^2})$ is negligibly small;(ii) the integral over $z$ become logarithmic when $z$ is much smaller than $r$. Applying the two rules, Eq.(\ref{ScaBK}) reduces to
\be
\frac{\del\bar{S}(r,\eta)}{\del\eta} \simeq -\abar\bar{S}(r,\eta)\int_{1/Q_s^2}^{r^2}\frac{\dif z^2}{z^2},
\label{SScaBK}
\ee
whose solution is
\be
\bar{S}(r,\eta) = \exp\left[-\frac{\abar^2}{2}\frac{\bar{\lambda}}{1+\abar\bar{\lambda}}\big(\eta - \eta_0\big)^2\right]\bar{S}(r,\eta_0),
\label{solcaBK}
\ee
where we have assumed the saturation momentum in $\eta$-representation as $\bar{Q}_s^2=Q_0^2\exp(\bar{\lambda}\eta)$. By comparing Eq.(\ref{solcaBK}) with Eq.(\ref{SolLOBK}), one can see that they have similar form, but the solution of the caBK-$\eta$ equation has an extra suppression factor in the exponent, which leads to the evolution speed of the dipole amplitude is slowed down. The numerical solutions of these two equations shall be done in the next section, where the numerical calculations support the aforementioned analytic result.

\subsection{Collinearly-improved BK equation in $\eta$: rcBK approximation in expansion of $S$-matrix}
In the previous subsection, the first derivative terms in the Taylor expansions in Eqs.(\ref{expandxz}) and (\ref{expandzy}) are approximately replaced by the LO BK equation, which are insufficient. To achieve the interested order of accuracy, we shall derive the collinearly-improved BK equation in $\eta$ by using the rcBK equation (\ref{rcBK}) to replace LO BK equation (\ref{LOBK}) in the expression of the first derivative terms.

\subsubsection{Extended caBK-$\eta$ equation}

Using the rcBK equation, one can re-expand the $S$-matrices in Eqs.(\ref{expandxz}) and (\ref{expandzy}) as

\bea
S_{\bx\bz}(Y)&=&\bar{S}_{\bx\bz} \left(\eta +\ln\frac{(\bx-\bz)^2}{(\bx-\by)^2}\right)\nn
&\simeq& \bar{S}_{\bx\bz}(\eta) + \ln\frac{(\bx-\bz)^2}{(\bx-\by)^2}
\frac{\partial \bar{S}_{\bx\bz}(\eta)}{\partial \eta}\nn
&\simeq& \bar{S}_{\bx\bz}(\eta) +\int\dif^2 \bu K^{\mathrm{rc}}(\bx,\bz,\bu)
\ln\frac{(\bx-\bz)^2}{(\bx-\by)^2}
\left[\bar{S}_{\bx\bu}(\eta) \bar{S}_{\bu\bz}(\eta) - \bar{S}_{\bx\bz}(\eta) \right],
\label{expandxzrc1}
\eea
and
\bea
S_{\bz\by}(Y)&=&\bar{S}_{\bz\by} \left(\eta +\ln\frac{(\by-\bz)^2}{(\bx-\by)^2}\right)\nn
&\simeq&\bar{S}_{\bz\by}(\eta)
+ \ln\frac{(\by-\bz)^2}{(\bx-\by)^2}
\frac{\partial \bar{S}_{\bx\bz}(\eta)}{\partial \eta}\nn
&\simeq&
\bar{S}_{\bz\by}(\eta)
+\int\dif^2 \bu K^{\mathrm{rc}}(\bz,\by,\bu)
\ln\frac{(\bz-\by)^2}{(\bx-\by)^2}
\left[\bar{S}_{\bz\bu}(\eta) \bar{S}_{\bu\by}(\eta) - \bar{S}_{\bz\by}(\eta) \right],
\label{expandzyrc1}
\eea
with the running coupling evolution kernel
\begin{align}
K^{\mathrm{rc}}(\bx,\by,\bz)=\frac{\abar}{2 \pi}
  \left[\frac{(\bx \minus \by)^2}{(\bx \minus \bz)^2\,(\by \minus \bz)^2}+
    \frac{1}{(\bx \minus \bz)^2}\left(\frac{\af^{xz}}{\af^{yz}}-1\right)+
    \frac{1}{(\by \minus \bz)^2}\left(\frac{\af^{yz}}{\af^{xz}}-1\right)
  \right],
\label{rcKernel}
\end{align}
where we use the shorthand notation $\af^{xz}=\af((\bx-\bz)^2)$ and similarly for others. Note that Eq.(\ref{rcKernel}) is another form of Eq.(\ref{Bal_rc}), expressed in terms of transverse coordinates of the dipoles.
Substituting the Eqs.(\ref{expandxzrc1}) and (\ref{expandzyrc1}) into Eq.(\ref{fnlobkY}), we obtain a semi-finished local collinearly-improved BK euqation in $\eta$ after some complicated algebra calculations (for the detailed derivation, see Appendix A),
\begin{align}
 \hspace*{0.5cm}
 \frac{\partial S_{\bx\by}(\eta)}{\partial \eta} &= \,
\frac{\abar}{2 \pi}
\int \dif^2 \bz\cdot K_0\cdot \,
\Big(\bar{S}_{\bx\bz}(\eta)\bar{S}_{\bz\by}(\eta)
\minus \bar{S}_{\bx\by}(\eta) \Big)
\nn
& \hspace*{0.5cm}
+\frac{\abar}{2 \pi}\int
\dif^2 \bz \cdot (K_q + K_g)\,
 \Big(\bar{S}_{\bx\bz}(\eta) \bar{S}_{\bz\by}(\eta) - \bar{S}_{\bx\by}(\eta) \Big)
\nn
& \hspace*{0.5cm}+ \frac{\abar^2}{2{\pi^2}} \int
	 \dif^2 \bz \,\dif^2 \bu\cdot K_{\mathrm{rc}} \cdot	
\bar{S}_{\bx\bu}(\eta)
 \Big(\bar{S}_{\bu\bz}(\eta) \bar{S}_{\bz\by}(\eta) - \bar{S}_{\bu\by}(\eta) \Big)
\nn
 & \hspace*{0.5cm} +\frac{\abar^2}{8\pi^2}
 \int\dif^2 \bu \,\dif^2 \bz\cdot K_1 \cdot
\Big(\bar{S}_{\bx\bu}(\eta) \bar{S}_{\bu\bz}(\eta) \bar{S}_{\bz\by}(\eta) - \bar{S}_{\bx \bu}(\eta) \bar{S}_{\bu \by}(\eta)\Big)
\nn
 & \hspace*{0.5cm} + \frac{\abar^2}{8\pi^2}\frac{N_f}{N_c}\int\dif^2 \bu \,\dif^2 \bz
\cdot K_f\cdot\Big(\bar{S}_{\bx\bz}(\eta)\bar{S}_{\bu\by}(\eta)-\bar{S}_{\bx\bu}(\eta)\bar{S}_{\bu\by}(\eta)\Big),
\label{fnlobkrc}
\end{align}
with
\begin{align}
 K_{\mathrm{rc}} &=\ln\frac{(\bu-\by)^2}{(\bx-\by)^2}\bigg[ \frac{(\bx \minus \by)^2}{(\bx \minus \bu)^2\,(\bu \minus \bz)^2\,(\by \minus \bz)^2}+
 \frac{(\bx \minus \by)^2}{(\bx \minus \bu)^2\,(\by \minus \bu)^2}\frac{1}{(\bu \minus \bz)^2}\left(\frac{\af^{uz}}{\af^{yz}}-1\right)
 \nn
  &\hspace*{1.8cm}+
 \frac{(\bx \minus \by)^2}{(\bx \minus \bu)^2\,(\by \minus \bu)^2(\by \minus \bz)^2}\left(\frac{\af^{yz}}{\af^{uz}}-1\right)+
 \frac{(\bu \minus \by)^2 }{(\bx \minus \bu)^2(\bu \minus \bz)^2\,(\by \minus \bz)^2}\left(\frac{\af^{xu}}{\af^{yu}}-1\right)
 \nn
  &\hspace*{1.8cm}+
 \frac{1}{(\bx \minus \bu)^2(\bu \minus \bz)^2}\left(\frac{\af^{xu}}{\af^{yu}}-1\right)\left(\frac{\af^{uz}}{\af^{yz}}-1\right)+
 \frac{1}{(\bx \minus \bu)^2(\by \minus \bz)^2}\left(\frac{\af^{xu}}{\af^{yu}}-1\right)\left(\frac{\af^{yz}}{\af^{uz}}-1\right)
 \nn
  &\hspace*{1.8cm}+
\frac{1}{(\bu \minus \bz)^2\,(\by \minus \bz)^2}\left(\frac{\af^{yu}}{\af^{xu}}-1\right)+
\frac{1}{(\by \minus \bu)^2(\bu \minus \bz)^2}\left(\frac{\af^{yu}}{\af^{xu}}-1\right)\left(\frac{\af^{uz}}{\af^{yz}}-1\right)
\nn
  &\hspace*{1.8cm}+
\frac{1}{(\by \minus \bu)^2(\by \minus \bz)^2}\left(\frac{\af^{yu}}{\af^{xu}}-1\right)\left(\frac{\af^{yz}}{\af^{uz}}-1\right)
 \bigg].
 \label{exrc}
\end{align}
By comparing Eq.(\ref{fnlobkrc}) with Eq.(\ref{fnlobk1}), there are extra eight terms resulting from the running coupling corrections, which have a significant impact on the evolution speed of the dipole amplitude. However, these extra terms do not cure the unstable issue of the evolution equation. Based on the discussion after Eq.(\ref{fnlobk1}), we know that the Eq.(\ref{fnlobkrc}) still has the instabilities caused by double collinear logarithms.

To curve the instability problem, the successive gluon emissions during the rapidity evolution must be ordered in lifetime and longitudinal momentum simultaneously. So, we need to put the constraints, Eqs.(\ref{ltcst}) and (\ref{lmcst}), on the successive gluon emissions as what we have done in the previous section,
\begin{align}
\bar S_{\bx\by}(\eta)= S^{(0)}_{\bx\by}+
\int\dif^2\bz K^{\mathrm{rc}}(\bx,\by,\bz)
\int\limits_
{\Theta(- \rho_1) |\rho_1|}^{\eta-\Theta(\rho_1- \rho)(\rho_1- \rho)}\dif \eta_1
\big[ \bar S_{\bx\bz}(\eta_1)
\bar S_{\bz\by}(\eta_1) \minus  \bar S_{\bx\by}(\eta_1) \big].
\label{BKetarc}
\end{align}
Note that originally, the integral form of the LO BK equation was used in the derivation of the ciBK-$\eta$ equation\cite{DIMST19}. In order to achieve the interested order of accuracy, we use the integral form of the rcBK equation instead of the LO BK equation as the start point to derive the ciBK-$\eta$ equation.
Performing derivatives over $\eta$ in Eq.(\ref{BKetarc}), one can get the exBK-$\eta$ equation
\begin{align}
\frac{\del \bar{S}_{\bx\by}(\eta)}{\del \eta} =
\int \dif^2 \bz K^{\mathrm{rc}}(\bx,\by,\bz)
\Theta\big(\eta \minus \delta_{\bx\by\bz}\big)
\big[\bar{S}_{\bx\bz}(\eta \minus \delta_{\bx\bz;r})\bar{S}_{\bz\by}(\eta \minus \delta_{\bz\by;r})
\minus \bar{S}_{\bx\by}(\eta) \big],
\label{edcaBK}
\end{align}
which has the same structure as the caBK-$\eta$ equation (\ref{caBK}), but with a running coupling modified kernel. In terms of the experience from $Y$-representation, we deduce that the rapidity evolution of the dipole amplitude is also suppressed by the modified kernel, which shall be approved by the numerical calculations in the next section.

\subsubsection{Extended full collinearly improved BK equation at $\mathcal{O}(\abar^2)$}
Based on the discussion in previous subsection, one can extend the exBK-$\eta$ equation to full collinearly-improved BK equation in $\eta$ by adding the NLO terms from Eq.(\ref{fnlobkrc}).
Before writing down the extended full collinearly-improved BK equation in $\eta$-representation, we need to identify the $\mathcal{O}(\abar^2)$ piece in the right hand side of Eq.(\ref{edcaBK}). First, we expand the $\bar{S}_{\bx\bz}$ and $\bar{S}_{\bz\by}$ to linear order in the rapidity shift, and then use the rcBK equation to replace the derivative terms,
\begin{align}
\bar{S}_{\bx\bz} \left(\eta - \delta_{\bx\bz;r}\right)
\simeq\,&
\bar{S}_{\bx\bz}(\eta)
- \delta_{\bx\bz;r}
\frac{\partial \bar{S}_{\bx\bz}(\eta)}{\partial \eta}
\nn
\simeq\, &
\bar{S}_{\bx\bz}(\eta)
-\int\dif^2 \bu K^{\mathrm{rc}}(\bx,\bz,\bu)
\delta_{\bx\bz;r}
\left[\bar{S}_{\bx\bu}(\eta) \bar{S}_{\bu\bz}(\eta) - \bar{S}_{\bx\bz}(\eta) \right],
\label{shiftexpandxznlrc}
\end{align}
and
\begin{align}
\bar{S}_{\bz\by} \left(\eta - \delta_{\bz\by;r}\right)
\simeq\,&
\bar{S}_{\bz\by}(\eta)
- \delta_{\bz\by;r}
\frac{\partial \bar{S}_{\bz\by}(\eta)}{\partial \eta}
\nn
\simeq\, &
\bar{S}_{\bz\by}(\eta)-\int\dif^2 \bu K^{\mathrm{rc}}(\bz,\by,\bu)
\delta_{\bz\by;r}
\left[\bar{S}_{\bz\bu}(\eta) \bar{S}_{\bu\by}(\eta) - \bar{S}_{\bz\by}(\eta) \right].
\label{shiftexpandzynlrc}
\end{align}
Substituting Eqs.(\ref{shiftexpandxznlrc}) and (\ref{shiftexpandzynlrc}) into the non-linear term in Eq.(\ref{edcaBK}), we get the $\mathcal{O}(\abar^2)$ piece in the right hand side of Eq.(\ref{edcaBK}),
\be
-2\int
	 \dif^2 \bz \,\dif^2 \bu K^{\mathrm{rc}}(\bx,\by,\bu)K^{\mathrm{rc}}(\bu,\by,\bz)
	\delta_{\bu\by;r}\
	\bar{S}_{\bx\bu}(\eta)
 \left[\bar{S}_{\bu\bz}(\eta) \bar{S}_{\bz\by}(\eta) - \bar{S}_{\bu\by}(\eta) \right],
\ee
where we have used the property that the running coupling terms are invariant under $\bx-\bz\rightarrow\bz-\by$ (for the detailed derivation, see Appendix B).

Now, we are arriving the final stage to get the extended evolution equation in $\eta$-representation. Subtracting the $\mathcal{O}(\abar^2)$ piece from Eq.(\ref{edcaBK}), and adding the $\mathcal{O}(\abar^2)$ pieces from Eq.(\ref{fnlobkrc}), we obtain the extended full collinearly-improved BK equation in $\eta$ as
\begin{align}
 \frac{\partial S_{\bx\by}(\eta)}{\partial \eta} &= \,
\frac{\abar}{2 \pi}
 \int\dif^2 \bz \cdot K_0\cdot\,
\Theta(\eta- \delta_{\bx\by\bz})\Big(\bar{S}_{\bx\bz}(\eta \minus \delta_{\bx\bz;r})\bar{S}_{\bz\by}(\eta \minus \delta_{\bz\by;r})
\minus \bar{S}_{\bx\by}(\eta)\Big)
\nn
& \hspace*{0.5cm}
+\frac{\abar}{2 \pi}\int
\dif^2 \bz \cdot (K_q + K_g)\,
 \Big(\bar{S}_{\bx\bz}(\eta) \bar{S}_{\bz\by}(\eta) - \bar{S}_{\bx\by}(\eta) \Big)
\nn
& \hspace*{0.5cm}+ \frac{\abar^2}{2{\pi^2}} \int
	 \dif^2 \bz \,\dif^2 \bu\cdot K_3 \cdot	
\bar{S}_{\bx\bu}(\eta)
 \Big(\bar{S}_{\bu\bz}(\eta) \bar{S}_{\bz\by}(\eta) - \bar{S}_{\bu\by}(\eta) \Big)
\nn
 & \hspace*{0.5cm} +\frac{\abar^2}{8\pi^2}
 \int\dif^2 \bu \,\dif^2 \bz\cdot K_1 \cdot
\Big(\bar{S}_{\bx\bu}(\eta) \bar{S}_{\bu\bz}(\eta) \bar{S}_{\bz\by}(\eta) - \bar{S}_{\bx \bu}(\eta) \bar{S}_{\bu \by}(\eta)\Big)
\nn
 & \hspace*{0.5cm} + \frac{\abar^2}{8\pi^2}\frac{N_f}{N_c}\int\dif^2 \bu \,\dif^2 \bz
\cdot K_f\cdot\Big(\bar{S}_{\bx\bz}(\eta)\bar{S}_{\bu\by}(\eta)-\bar{S}_{\bx\bu}(\eta)\bar{S}_{\bu\by}(\eta)\Big),
\label{maineq}
\end{align}
with
\be
K_3 = K_{\mathrm{rc}} + \delta_{\bu\by;r},
\ee
which is non-local in $\eta$, and a stable equation. The double anti-collinear logarithmic term in the second line in the right hand side of Eq.(\ref{maineq}) is canceled by the relevant piece generated via the integral over $\bu$ in the third term. The Eq.(\ref{maineq}) is free of double collinear logarithms, since all these logarithms are fully included in the first term.
By comparing Eq.(\ref{maineq}) with Eq.(\ref{fnlobk}), one can find that in the right hand side of Eq.(\ref{maineq}) there are eight extra terms which are resulting from the running coupling corrections in the Taylor expansion of the $S$-matrix. These extra terms play a significant role in the suppression of the evolution speed of the dipole amplitude.

\subsubsection{Analytic solution to the exBK-$\eta$ equation in saturation region}
\label{sec:solexBK}
Let us move to analytically solve the exBK-$\eta$ equation, Eq.(\ref{edcaBK}), in saturation region. In this regime, we know that one of the two daughter dipoles has similar size as the parent dipole, but the size of the rest one is much smaller than the parent dipole. Moreover, it is known that the non-locality is only important for the $S$-matrix which is associated with small size. Therefore, we can reduce Eq.(\ref{edcaBK}) to
\begin{align}
\frac{\del \bar{S}(r,\eta)}{\del \eta} \simeq
2\frac{1}{2\pi}\bar{S}(r,\eta)\int_{1/Q_s}^{r} \frac{\dif^2z}{z^2}\abar(z^2)
\Theta\Big(\eta \minus \ln\frac{r^2}{z^2}\Big)
\left[\bar{S}\Big(z, \eta \minus \ln\frac{r^2}{z^2}\Big)
\minus 1\right],
\label{edcaBKS}
\end{align}
where the factor 2 accounts for the smaller size dipole coming from any one of the two daughter dipoles, and the smallest size of the dipoles ($z$) is used to be as the argument of the QCD coupling.

To solve Eq.({\ref{edcaBKS}}) analytically in saturation region, the same strategy is used as what we have done in Sec.\ref{sec:solcaBK}. The one loop running coupling, Eq.(\ref{oneloop}), is used in the calculations. The Eq.({\ref{edcaBKS}}) becomes
\begin{align}
\frac{\del \bar{S}(r,\eta)}{\del \eta} \simeq
-\frac{N_c}{\pi}\bar{S}(r,\eta)\int_{1/Q_s^2}^{r^2} \frac{\dif z^2}{z^2}\frac{1}{b\ln\frac{1}{z^2\Lambda^2}}.
\label{edcaBKSS}
\end{align}
Performing the integral over $z$, we get
\be
\frac{\del \bar{S}(r,\eta)}{\del \eta} \simeq -\frac{N_c}{b\pi}\lt[\ln\lt(\ln\frac{Q_s^2}{\Lambda^2}\rt)-\ln\lt(\ln\frac{1}{r^2\Lambda^2}\rt)\rt],
\ee
whose solution is
\be
\bar{S}(r,\eta) = \exp\lt\{-\frac{N_c}{b\pi}(\eta - \eta_0)\lt[\ln\lt(\frac{\sqrt{\bar{\lambda'}}(\eta-\eta_0)+\frac{\sqrt{\bar{\lambda'}}}{2}\ln\frac{1}{r^2\Lambda^2}}{(\sqrt{\eta-\eta_0}+\frac{\sqrt{\bar{\lambda}'}}{2})\ln\frac{1}{r^2\Lambda^2}}\rt)-\frac{1}{2}\rt]\rt\}\bar{S}(r,\eta_0),
\label{solexBK}
\ee
with the saturation momentum in NLO case,
\be
\ln\frac{\bar{Q}_s^2}{\Lambda^2}=\sqrt{\bar{\lambda}'(\eta-\eta_0)} + \mathcal{O}(\eta^{1/6}),
\ee
and
\be
r^2Q_s^2 \simeq \lt[r^2\bar{Q}_s^2\rt]^\frac{1}{1+\sqrt{\frac{\bar{\lambda}'}{4\eta}}}.
\ee
By comparing Eq.(\ref{solexBK}) with Eq.(\ref{solcaBK}), one can see that the quadratic rapidity dependence in the exponent of the $S$-matrix is replaced by the linear rapidity dependence once the running coupling corrections are taken into account. This change implies that the evolution speed of the dipole amplitude is suppressed by the NLO corrections.

\section{Numerical analysis}
\label{sec:numsol}
To test the analytic results obtained in the above section, we shall numerically solve the evolution equations in this section. The Eqs.(\ref{caBK}) and (\ref{edcaBK}) are integro-differential equations which can be numerically straightforward solved on a lattice. To simplify the computation, we neglect impact parameter dependence of the dipole amplitude throughout this numerical calculations, which imply that the dipole amplitude does not depend on angle, $N(\br, Y)= 1-S(\br, Y)=1-S(|r|, Y)$. Thus, we can view the evolution equations as a set of differential equations and solve them at discrete values of transverse separation. To be more specific, we discretize the dipole transverse size $r$ into 800 points which are equally located in the logarithmic space between $r_{\mathrm{min}}=10^{-8}\mathrm{GeV}^{-1}$ and $r_{\mathrm{max}}=50\mathrm{GeV}^{-1}$.
The GNU Scientific Library (GSL) is a good candidate to solve them, since the GSL contains almost all the routines required by our purpose, such as the Runge-Kutta method for solving differential equations, adaptive integral routines for performing numerical integrals, and the cubic spline interpolation codes for interpolating the data points not located on the lattice.

The initial condition for the evolution equations is parameterized at rapidity $\eta=0$. We use the Golec-Biernat and Wusthoff (GBW) parametrization as the initial condition\cite{GBW},
\be
N^{\mathrm{GBW}}(r, \eta=0) = \lt\{1-\exp\lt[-\lt(\frac{r^2 Q_{\mathrm{s0}}^2}{4}\rt)^p\rt]\rt\}^{1/p},
\label{IC}
\ee
with $p=4$, and $Q_{\mathrm{s0}}=0.362\mathrm{GeV}$\cite{DIST20}.

For the strong coupling constant, we use the one-loop running coupling, Eq.(\ref{oneloop}), with $N_f=3$ and $N_c=3$. According to the performance of the running coupling in the fit to HERA data\cite{IMMST2,Xiang21}, we choose to use the smallest dipole running coupling prescription which means the argument of coupling is the smallest dipole among the parent and daughter dipoles,
\be
\af(r_{\mathrm{min}}^2) = \af\big(\mathrm{min}\{(\bx-\by)^2,(\bx-\bz)^2,(\bz-\by)^2\}\big).
\ee
We freeze the running coupling $\alpha_s(r_{\mathrm{fr}})=0.75$ when $r>r_{\mathrm{fr}}$ in order to regularize the infrared behavior.

To show the impact of the running coupling corrections on the speed of the fronts, the saturation exponent is numerically calculated
\be
\bar{\lambda} = \frac{\mathrm{d}\ln \bar{Q}_s^2(\eta)}{\mathrm{d}\eta},
\ee
where the saturation moment $\bar{Q}_s(\eta)$ is determined by $N(r=1/\bar{Q}_s,\eta)=\kappa$ with $\kappa$ to be a constant of order 1.

\begin{figure}[h!]
\setlength{\unitlength}{1.5cm}
\begin{center}
\epsfig{file=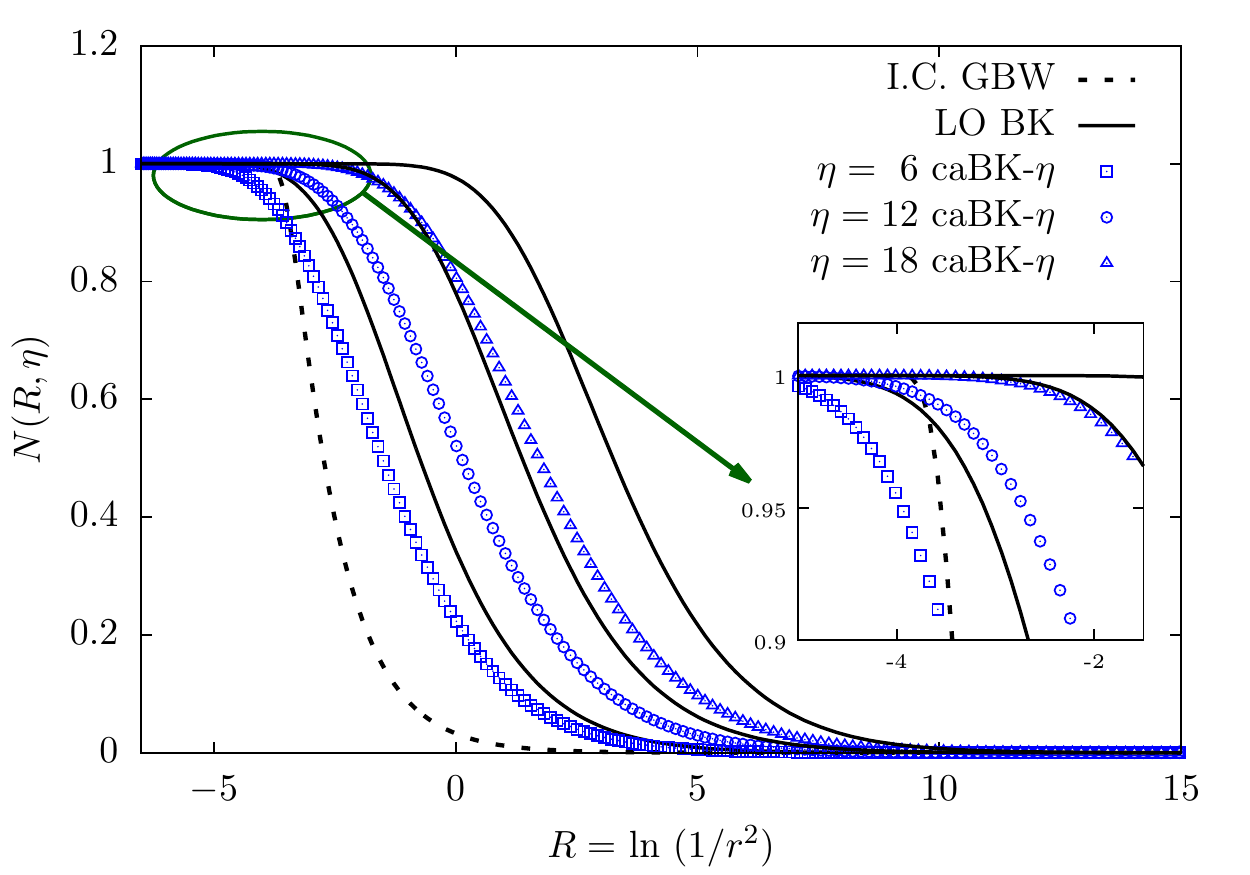, width=8cm,height=6cm}
\epsfig{file=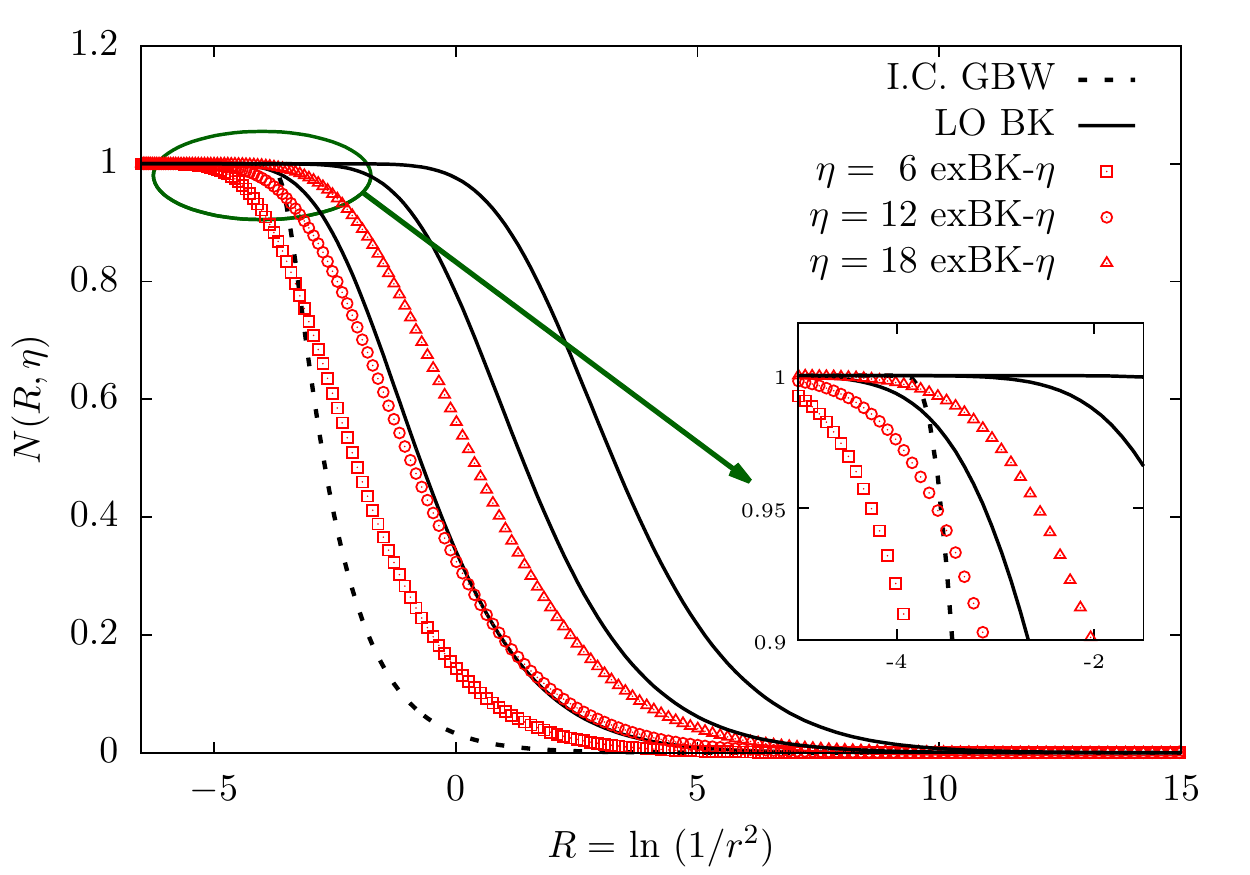, width=8cm,height=6cm}
\end{center}
\caption{The numerical solutions to LO BK, caBK-$\eta$, and exBK-$\eta$ equations for 4 different rapidities. The left hand panel shows the comparisons of the evolution speed between the LO BK and caBK-$\eta$ dipole amplitudes. The right hand panel gives the comparisons of the evolution speed between LO BK and exBK-$\eta$ dipole amplitudes. The zooming diagrams show the relevant results in the saturation region.}
\label{figBK}
\end{figure}

The left-hand panel of Fig.\ref{figBK} gives the solutions of the LO BK and caBK-$\eta$ equations for 4 different rapidities. By comparing the solutions for each respective rapidities, one can see that the values of the caBK-$\eta$ dipole amplitude are smaller than the LO BK ones, which indicate that the NLO corrections enhanced by the double transverse logarithms suppress the evolution speed of the dipole amplitude. A zooming in diagram is provided to clearly show the numerical results in the saturation region. One can see that the evolution is also slowed down in the saturation region. This numerical outcome is consistent with the analytic results in Eqs.(\ref{SolLOBK}) and (\ref{solcaBK}), where the analytic solutions of the LO BK and caBK-$\eta$ have analogous expression, but with different coefficients in the exponent. The coefficient in the caBK-$\eta$ case is smaller than the LO BK one, which leads to the caBK-$\eta$ dipole amplitude smaller than LO BK one. The right-hand panel of Fig.\ref{figBK} shows the comparison of the solutions of LO BK and exBK-$\eta$ equation for 4 different rapidities. We plot a inner zooming in diagram for a clear comparison between the LO BK and exBK-$\eta$ dipole amplitudes in saturation region. One can see, from the zooming in diagram, that the respective solution of the exBK-$\eta$ equation in each rapidity is much smaller than the LO BK one, which implies that the evolution speed of the dipole amplitude is significantly suppressed. The suppression is much more than the caBK-$\eta$ case. This outcome confirms the analytic result in Sec.\ref{sec:solexBK}, where the dipole amplitude is suppressed by the running coupling corrections.

\begin{figure}[h!]
\setlength{\unitlength}{1.5cm}
\begin{center}
\epsfig{file=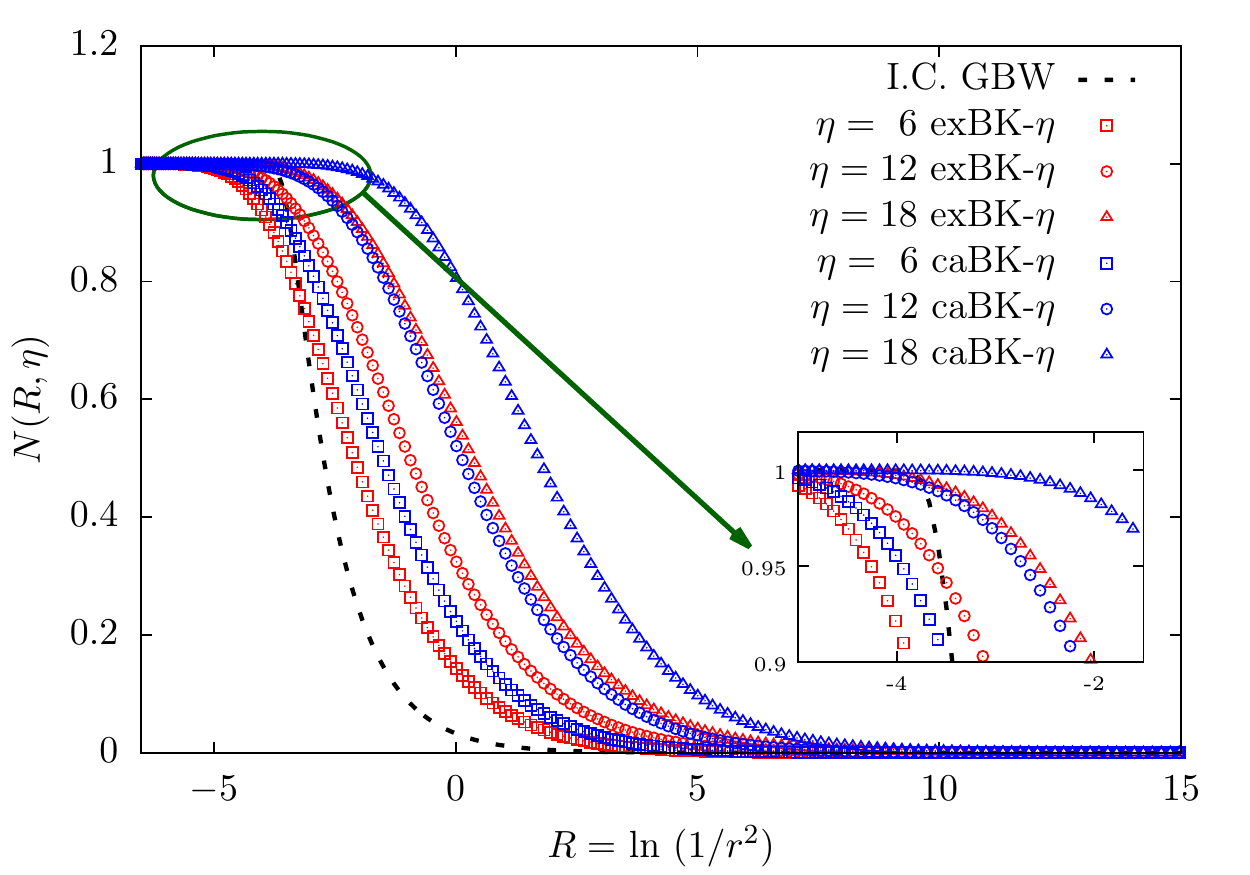, width=8cm,height=6cm}
\epsfig{file=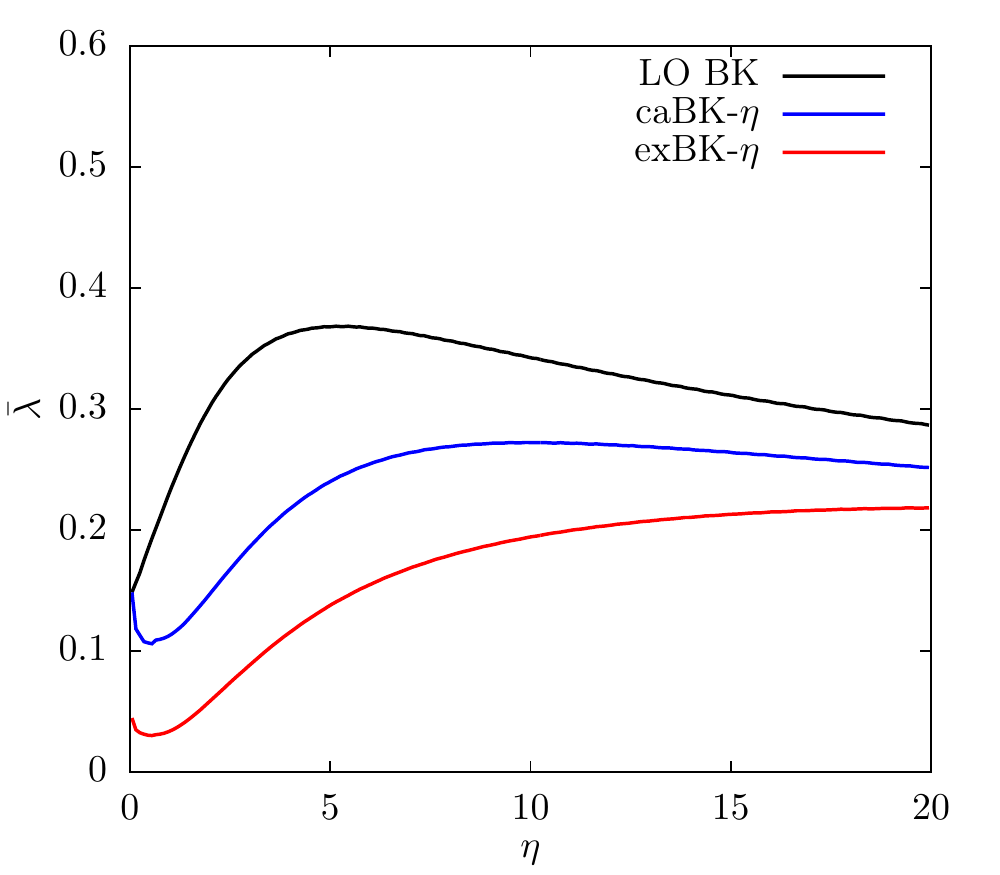, width=8cm,height=6cm}
\end{center}
\caption{The left hand panel gives the comparisons of the evolution speed between the caBK-$\eta$ and exBK-$\eta$ dipole amplitudes. The inner diagrams are the zooming in amplitudes in the saturation region. The right hand panel shows the saturation exponent as a function of $\eta$ in the LO BK, caBK-$\eta$, and exBK-$\eta$ cases.}
\label{figSM}
\end{figure}

The left-hand panel of Fig.\ref{figSM} gives the comparisons of the solutions of the caBK-$\eta$ and exBK-$\eta$ equations for 4 different rapidities. We can see that the respective dipole amplitudes resulting from the exBK-$\eta$ equation are smaller than the ones from caBK-$\eta$ equation. Especially, it clearly show the suppression in the saturation region from the inner zooming in diagram. This outcome supports the analytic findings in Eq.(\ref{solexBK}), where the dipole amplitude is further suppressed by the running coupling corrections on top of the double logarithm resummation. Finally, we present the $\eta$ dependence of the saturation exponent predicted by the LO BK, caBK-$\eta$, and exBK-$\eta$ in the right-hand panel of Fig.\ref{figSM}. As expected, the $\bar{\lambda}$ resulting from exBK-$\eta$ equation is the smallest one among them, which is consistent with the theoretical expectations, since the exBK-$\eta$ equation includes the running coupling corrections on top of the collinear resummations.

\begin{acknowledgments}
This work is supported by the National Natural Science Foundation of China under Grant Nos.11765005, 11305040, 11947119 and 11847152; the Fund of Science and Technology Department of Guizhou Province under Grant Nos.[2018]1023, and [2019]5653; the Education Department of Guizhou Province under Grant No.KY[2017]004; the National Key Research and Development Program of China under Grant No.2018YFE0104700, and Grant No.CCNU18ZDPY04.
\end{acknowledgments}

\appendix

\section{Local collinearly-improved BK equation in $\eta$ with running coupling corrections in Taylor expansion}

We give a detailed derivation of the collinearly-improved BK equation in $\eta$-representation by using the rcBK equation to expand out $S_{\bx\bz}(Y)$ in Eq.(\ref{expandxzrc1}) and $S_{\bz\by}(Y)$ in Eq.(\ref{expandzyrc1}). To simplify the calculations, we rewrite the running coupling evolution kernel, Eq.(\ref{rcKernel}), as
\begin{align}
\label{A1}
K^{\mathrm{rc}}(\bx,\by,\bz)=\frac{\abar}{2 \pi}(K_0+K_{q'}),
\end{align}
with
\be
K_{q'} = \frac{1}{(\bx \minus \bz)^2}\left(\frac{\af^{xz}}{\af^{yz}}-1\right)+
    \frac{1}{(\by \minus \bz)^2}\left(\frac{\af^{yz}}{\af^{xz}}-1\right).
\ee
The full NLO BK equation, Eq.(\ref{fnlobkY}), is our starting point of this derivation. Using Eq.(\ref{A1}), Eq.(\ref{fnlobkY}) is rewritten as
\begin{align}
 \label{A2}
 \frac{\partial S_{\bx\by}(Y)}{\partial Y} =& \,
 \int
\dif^2 \bz \, K^{\mathrm{rc}}(\bx,\by,\bz) \big[S_{\bx\bz}(Y) S_{\bz\by}(Y) - S_{\bx\by}(Y) \big]
\nn
 &  +\frac{\abar}{2\pi} \int \dif^2 \bz \cdot K_g \cdot \left[S_{\bx\bz}(Y) S_{\bz\by}(Y) - S_{\bx\by}(Y) \right]
 \nn
  &  +
\frac{\abar^2}{8\pi^2}
 \int \dif^2 \bu \,\dif^2 \bz \cdot K_1 \cdot \left[S_{\bx\bu}(Y) S_{\bu\bz}(Y) S_{\bz\by}(Y) - S_{\bx \bu}(Y) S_{\bu \by}(Y)\right]
\nn
  &  +
\frac{\abar^2}{8\pi^2}\frac{N_f}{N_c}
 \int \dif^2 \bu \,\dif^2 \bz  \cdot K_f \cdot \left[S_{\bx\bz}(Y) S_{\bu\by}(Y) - S_{\bx \bu}(Y) S_{\bu \by}(Y)\right].
 \end{align}
Note that the kernel $K^{\mathrm{rc}}$ is in the order of $\mathcal{O}(\abar)$, which means the expansions of $S_{\bx\bz}(Y)$ and $S_{\bz\by}(Y)$ in Eqs.(\ref{expandxzrc1}) and (\ref{expandzyrc1}) are equivalent to adding a term of order $\mathcal{O}(\abar)$. So we can deduce the rapidity shift by using a similar scheme in LO expansion case. For the first term (to be denoted as $T_{\mathrm{rc}}$) in the right hand side of Eq.(\ref{A2}), we have
\begin{align}
 \label{A3}
T_{\mathrm{rc}} &=\int \dif^2 \bz \, K^{\mathrm{rc}}(\bx,\by,\bz) \big[S_{\bx\bz}(Y) S_{\bz\by}(Y) - S_{\bx\by}(Y) \big]
\nn
 &= \,
  \int
\dif^2 \bz \, K^{\mathrm{rc}}(\bx,\by,\bz) \bigg[\bar{S}_{\bx\bz}\left(\eta +\ln\frac{(\bx-\bz)^2}{(\bx-\by)^2}\right) \bar{S}_{\bz\by}\left(\eta +\ln\frac{(\by-\bz)^2}{(\bx-\by)^2}\right) - \bar{S}_{\bx\by}(\eta) \bigg]
.
 \end{align}
Substituting Eqs.(\ref{expandxzrc1}) and (\ref{expandzyrc1}) into Eq.(\ref{A3}), one can get
 \begin{align}
 \label{A4}
T_{\mathrm{rc}}  & = \,
  \int
 \dif^2 \bz \, K^{\mathrm{rc}}(\bx,\by,\bz)
\left\{\left[\bar{S}_{\bx\bz}(\eta)
	+
		\int
	\dif^2 \bu K^{\mathrm{rc}}(\bx,\bz,\bu)
	\ln\frac{(\bx-\bz)^2}{(\bx-\by)^2}
	\left[\bar{S}_{\bx\bu}(\eta) \bar{S}_{\bu\bz}(\eta) - \bar{S}_{\bx\bz}(\eta) \right]\right] \right.
\nn
 &\hspace*{1.0cm} \times \left.\left[\bar{S}_{\bz\by}(\eta)
	+ \int
	\dif^2 \bu K^{\mathrm{rc}}(\bz,\by,\bu)
	\ln\frac{(\bz-\by)^2}{(\bx-\by)^2}
	\left[\bar{S}_{\bz\bu}(\eta) \bar{S}_{\bu\by}(\eta) - \bar{S}_{\bz\by}(\eta) \right]\right]
 - \bar{S}_{\bx\by}(\eta) \right\}.
 \end{align}
After some algebra calculations, Eq.(\ref{A4}) becomes
 \begin{align}
 \label{A5}
T_{\mathrm{rc}}  & = \,
 \int
 \dif^2 \bz \, K^{\mathrm{rc}}(\bx,\by,\bz)
\Big\{\left[\bar{S}_{\bx\bz}(\eta)\bar{S}_{\bz\by}(\eta)- \bar{S}_{\bx\by}(\eta)\right]
+
	\bar{S}_{\bx\bz}(\eta)\int
	\dif^2 \bu K^{\mathrm{rc}}(\bz,\by,\bu)
	\ln\frac{(\bz-\by)^2}{(\bx-\by)^2}
\nn
&	
	\hspace*{1.0cm}  \times \left[\bar{S}_{\bz\bu}(\eta) \bar{S}_{\bu\by}(\eta) - \bar{S}_{\bz\by}(\eta) \right]
+	
	\bar{S}_{\bz\by}(\eta)\int
	\dif^2 \bu K^{\mathrm{rc}}(\bx,\bz,\bu)
	\ln\frac{(\bx-\bz)^2}{(\bx-\by)^2}
	\left[\bar{S}_{\bx\bu}(\eta) \bar{S}_{\bu\bz}(\eta) - \bar{S}_{\bx\bz}(\eta) \right]
\nn
 &	\hspace*{1.0cm} +\int
	\dif^2 \bu \dif^2 \bv K^{\mathrm{rc}}(\bx,\bz,\bu)K^{\mathrm{rc}}(\bz,\by,\bv)
	\ln\frac{(\bx-\bz)^2}{(\bx-\by)^2}\ln\frac{(\bz-\by)^2}{(\bx-\by)^2}
\left[\bar{S}_{\bx\bu}(\eta) \bar{S}_{\bu\bz}(\eta) - \bar{S}_{\bx\bz}(\eta) \right]
\nn
 &	\hspace*{1.0cm} \times
 \left[\bar{S}_{\bz\bv}(\eta) \bar{S}_{\bv\by}(\eta) - \bar{S}_{\bz\by}(\eta) \right]
  \Big\}
 .
 \end{align}

Using the property that the running coupling terms are invariant under $\bx\minus\bz \to \bz\minus\by$, we can see that the second term is equal to the third term in the brace of Eq.(\ref{A5}). In addition, the fourth term in the brace of Eq.(\ref{A5}) is in the order of $\mathcal{O}(\abar^2)$. It becomes $\mathcal{O}(\abar^3)$ order because of an extra $K^{\mathrm{rc}}$ factor in front the brace and can be discarded. Therefore, Eq.(\ref{A5}) can be reduced to
\begin{align}
\label{A6}
 T_{\mathrm{rc}}  =& \,
  \int
 \dif^2 \bz \, K^{\mathrm{rc}}(\bx,\by,\bz)
\left[\bar{S}_{\bx\bz}(\eta)\bar{S}_{\bz\by}(\eta)- \bar{S}_{\bx\by}(\eta)\right]
\nn*[0.2cm]
&	+
2\int
	 \dif^2 \bz \,\dif^2 \bu K^{\mathrm{rc}}(\bx,\by,\bz)K^{\mathrm{rc}}(\bz,\by,\bu)
	\ln\frac{(\bz-\by)^2}{(\bx-\by)^2}
	\bar{S}_{\bx\bz}(\eta)\left[\bar{S}_{\bz\bu}(\eta) \bar{S}_{\bu\by}(\eta) - \bar{S}_{\bz\by}(\eta) \right]
 .
 \end{align}
In order to keep consistence with the physics picture, we relabel the integral variables in the second term in the right hand side of Eq.(\ref{A6}) through variable transformation $\bu \leftrightarrow \bz$. Then Eq.(\ref{A6}) becomes
\begin{align}
\label{A7}
 T_{\mathrm{rc}} = &\,
  \int
 \dif^2 \bz \, K^{\mathrm{rc}}(\bx,\by,\bz)
\left[\bar{S}_{\bx\bz}(\eta)\bar{S}_{\bz\by}(\eta)- \bar{S}_{\bx\by}(\eta)\right]
\nn*[0.2cm]
&	+
2\int
	 \dif^2 \bz \,\dif^2 \bu K^{\mathrm{rc}}(\bx,\by,\bu)K^{\mathrm{rc}}(\bu,\by,\bz)
	\ln\frac{(\bu-\by)^2}{(\bx-\by)^2}
	\bar{S}_{\bx\bu}(\eta)
 \left[\bar{S}_{\bu\bz}(\eta) \bar{S}_{\bz\by}(\eta) - \bar{S}_{\bu\by}(\eta) \right]
 .
 \end{align}

The last three terms in the right hand side of Eq.(\ref{A2}) are in order of $\mathcal{O}(\abar^2)$. For these three terms, we can simply replace the rapidity shift, like $S_{\bx \bu}(Y)\rightarrow S_{\bx \bu}(\eta)$, since the rapidity shift makes them to be of order $\mathcal{O}(\abar^3)$ which can be safely neglected in our study. Combining these terms with $T_{\mathrm{rc}}$ in Eq.(\ref{A7}), one can obtain
\begin{align}
 \label{A8}
  \frac{\partial \bar{S}_{\bx\by}(\eta)}{\partial \eta} = & \,
 \int
 \dif^2 \bz \, K^{\mathrm{rc}}(\bx,\by,\bz)
\left[\bar{S}_{\bx\bz}(\eta)\bar{S}_{\bz\by}(\eta)- \bar{S}_{\bx\by}(\eta)\right]
\nn
&	+
2\int
	 \dif^2 \bz \,\dif^2 \bu K^{\mathrm{rc}}(\bx,\by,\bu)K^{\mathrm{rc}}(\bu,\by,\bz)
	\ln\frac{(\bu-\by)^2}{(\bx-\by)^2}
	\bar{S}_{\bx\bu}(\eta)
 \left[\bar{S}_{\bu\bz}(\eta) \bar{S}_{\bz\by}(\eta) - \bar{S}_{\bu\by}(\eta) \right]
\nn
 &  +\frac{\abar}{2\pi} \int \dif^2 \bz \cdot K_g \cdot \left[\bar{S}_{\bx\bz}(\eta) \bar{S}_{\bz\by}(\eta) - \bar{S}_{\bx\by}(\eta) \right]
 \nn
  &  +
\frac{\abar^2}{8\pi^2}
 \int \dif^2 \bu \,\dif^2 \bz \cdot K_1 \cdot \left[\bar{S}_{\bx\bu}(\eta) \bar{S}_{\bu\bz}(\eta) \bar{S}_{\bz\by}(\eta) - \bar{S}_{\bx \bu}(\eta) \bar{S}_{\bu \by}(\eta)\right]
\nn
  &  +
\frac{\abar^2}{8\pi^2}\frac{N_f}{N_c}
 \int \dif^2 \bu \,\dif^2 \bz  \cdot K_f \cdot  \left[\bar{S}_{\bx\bz}(\eta) \bar{S}_{\bu\by}(\eta)  - \bar{S}_{\bx \bu}(\eta) \bar{S}_{\bu \by}(\eta)\right].
 \end{align}

The first integral term (to be denoted as $T_\mathrm{LK}$) and second integral term (to be denoted as $T_\mathrm{SK}$) in Eq.(\ref{A8}) contain linear $K^{\mathrm{rc}}$ factor and quadratic $K^{\mathrm{rc}}$ factor, respectively. For a clear comparison with the LO expansion case, we expend out these two terms by using Eq.(\ref{A1})
\begin{align}
 \label{A9}
T_\mathrm{LK} & = \,
\int
 \dif^2 \bz \, K^{\mathrm{rc}}(\bx,\by,\bz)
\left[\bar{S}_{\bx\bz}(\eta)\bar{S}_{\bz\by}(\eta)- \bar{S}_{\bx\by}(\eta)\right]
\nn
&=   \frac{\abar}{2\pi} \int \dif^2 \bz \cdot K_0 \cdot \left[\bar{S}_{\bx\bz}(\eta) \bar{S}_{\bz\by}(\eta) - \bar{S}_{\bx\by}(\eta) \right]
+\frac{\abar}{2\pi} \int \dif^2 \bz \cdot K_q \cdot \left[\bar{S}_{\bx\bz}(\eta) \bar{S}_{\bz\by}(\eta) - \bar{S}_{\bx\by}(\eta) \right],
 \end{align}
and
\begin{align}
 \label{A10}
 T_\mathrm{SK} =& \,2\int
	 \dif^2 \bz \,\dif^2 \bu K^{\mathrm{rc}}(\bx,\by,\bu)K^{\mathrm{rc}}(\bu,\by,\bz)
	\ln\frac{(\bu-\by)^2}{(\bx-\by)^2}
	\bar{S}_{\bx\bu}(\eta)
 \left[\bar{S}_{\bu\bz}(\eta) \bar{S}_{\bz\by}(\eta) - \bar{S}_{\bu\by}(\eta) \right]
   \nn
   =&
2\int
	 \dif^2 \bz \,\dif^2 \bu \frac{\abar}{2 \pi}
  \left[\frac{(\bx \minus \by)^2}{(\bx \minus \bu)^2\,(\by \minus \bu)^2}+
    \frac{1}{(\bx \minus \bu)^2}\left(\frac{\af^{xu}}{\af^{yu}}-1\right)+
    \frac{1}{(\by \minus \bu)^2}\left(\frac{\af^{yu}}{\af^{xu}}-1\right)
  \right]
  \nn
  &\hspace*{1.5cm} \times \frac{\abar}{2 \pi}\left[\frac{(\bu \minus \by)^2}{(\bu \minus \bz)^2\,(\by \minus \bz)^2}+
    \frac{1}{(\bu \minus \bz)^2}\left(\frac{\af^{uz}}{\af^{yz}}-1\right)+
    \frac{1}{(\by \minus \bz)^2}\left(\frac{\af^{yz}}{\af^{uz}}-1\right)
  \right]
  \nn
  &\hspace*{1.5cm} \times
	\ln\frac{(\bu-\by)^2}{(\bx-\by)^2}
	\bar{S}_{\bx\bu}(\eta)
 \left[\bar{S}_{\bu\bz}(\eta) \bar{S}_{\bz\by}(\eta) - \bar{S}_{\bu\by}(\eta) \right]
 .
 \end{align}
After some algebra calculations, Eq.(\ref{A10}) becomes
\begin{align}
 \label{A11}
 T_\mathrm{SK} = \,
\frac{\abar^2}{2{\pi^2}} \int
	 \dif^2 \bz \,\dif^2 \bu \cdot K_\mathrm{rc} \cdot
	\bar{S}_{\bx\bu}(\eta)
 \left[\bar{S}_{\bu\bz}(\eta) \bar{S}_{\bz\by}(\eta) - \bar{S}_{\bu\by}(\eta) \right],
 \end{align}
where the $K_\mathrm{rc}$ is defined in Eq.(\ref{exrc}).

Substituting the Eqs.(\ref{A9}) and (\ref{A11}) into Eq.(\ref{A8}), one can obtain a semi-finished collinearly-improved BK equation in $\eta$
\begin{align}
 \label{A12}
  \frac{\partial \bar{S}_{\bx\by}(\eta)}{\partial \eta} = & \,
 \frac{\abar}{2\pi} \int
 \dif^2 \bz \, \cdot K_0 \cdot
\left[\bar{S}_{\bx\bz}(\eta)\bar{S}_{\bz\by}(\eta)- \bar{S}_{\bx\by}(\eta)\right]
\nn
 &  +\frac{\abar}{2\pi} \int \dif^2 \bz \cdot (K_q +K_g) \cdot \left[\bar{S}_{\bx\bz}(\eta) \bar{S}_{\bz\by}(\eta) - \bar{S}_{\bx\by}(\eta) \right]
 \nn
  &  +
 \frac{\abar^2}{2{\pi^2}} \int
	 \dif^2 \bz \,\dif^2 \bu \cdot K_\mathrm{rc} \cdot
	\bar{S}_{\bx\bu}(\eta)
 \left[\bar{S}_{\bu\bz}(\eta) \bar{S}_{\bz\by}(\eta) - \bar{S}_{\bu\by}(\eta) \right]
  \nn
  &  +
\frac{\abar^2}{8\pi^2}
 \int \dif^2 \bu \,\dif^2 \bz \cdot K_1 \cdot \left[\bar{S}_{\bx\bu}(\eta) \bar{S}_{\bu\bz}(\eta) \bar{S}_{\bz\by}(\eta) - \bar{S}_{\bx \bu}(\eta) \bar{S}_{\bu \by}(\eta)\right]
\nn
  &  +
\frac{\abar^2}{8\pi^2}\frac{N_f}{N_c}
 \int \dif^2 \bu \,\dif^2 \bz  \cdot K_f \cdot  \left[\bar{S}_{\bx\bz}(\eta) \bar{S}_{\bu\by}(\eta) - \bar{S}_{\bx \bu}(\eta) \bar{S}_{\bu \by}(\eta)\right].
 \end{align}

\section{Identify an $\mathcal{O}(\abar^2)$ piece in Eq.(\ref{edcaBK})}

In this appendix, we give the details of the derivation to identify an $\mathcal{O}(\abar^2)$ piece in the right hand side of Eq.(\ref{edcaBK}).
Substituting Eqs.(\ref{shiftexpandxznlrc}) and (\ref{shiftexpandzynlrc}) into Eq.(\ref{edcaBK}) and neglecting the step function in it, one can get
 \begin{align}
 \label{B1}
\frac{\partial \bar{S}_{\bx\by}(\eta)}{\partial \eta}& = \,
  \int
 \dif^2 \bz \, K^{\mathrm{rc}}(\bx,\by,\bz)
\left\{\left[\bar{S}_{\bx\bz}(\eta)
	-
		\int
	\dif^2 \bu K^{\mathrm{rc}}(\bx,\bz,\bu)
	\delta_{\bx\bz;r}
	\left[\bar{S}_{\bx\bu}(\eta) \bar{S}_{\bu\bz}(\eta) - \bar{S}_{\bx\bz}(\eta) \right]\right] \right.
\nn
 &\hspace*{1.0cm} \times \left.\left[\bar{S}_{\bz\by}(\eta)
	-
		\int
	\dif^2 \bu K^{\mathrm{rc}}(\bz,\by,\bu)
	\delta_{\bz\by;r}
	\left[\bar{S}_{\bz\bu}(\eta) \bar{S}_{\bu\by}(\eta) - \bar{S}_{\bz\by}(\eta) \right]\right]
 - \bar{S}_{\bx\by}(\eta) \right\}
  .
 \end{align}
Expanding the terms in the brace in the right hand side of Eq.(\ref{B1}), we obtain
  \begin{align}
 \label{B2}
 \frac{\partial \bar{S}_{\bx\by}(\eta)}{\partial \eta}& = \,
  \int
 \dif^2 \bz \, K^{\mathrm{rc}}(\bx,\by,\bz)
\Big\{\left[\bar{S}_{\bx\bz}(\eta)\bar{S}_{\bz\by}(\eta)- \bar{S}_{\bx\by}(\eta)\right]
-
	\bar{S}_{\bx\bz}(\eta)\int
	\dif^2 \bu K^{\mathrm{rc}}(\bz,\by,\bu)
	\delta_{\bz\by;r}
\nn
 &\hspace*{1.0cm} \times \left[\bar{S}_{\bz\bu}(\eta) \bar{S}_{\bu\by}(\eta) - \bar{S}_{\bz\by}(\eta) \right]
-	
	\bar{S}_{\bz\by}(\eta)\int
	\dif^2 \bu K^{\mathrm{rc}}(\bx,\bz,\bu)
	\delta_{\bx\bz;r}
	\left[\bar{S}_{\bx\bu}(\eta) \bar{S}_{\bu\bz}(\eta) - \bar{S}_{\bx\bz}(\eta) \right]
\nn
 &+\int
	\dif^2 \bu \dif^2 \bv K^{\mathrm{rc}}(\bx,\bz,\bu)K^{\mathrm{rc}}(\bz,\by,\bv)
	\delta_{\bx\bz;r}\delta_{\bz\by;r}
 \left[\bar{S}_{\bx\bu}(\eta) \bar{S}_{\bu\bz}(\eta) - \bar{S}_{\bx\bz}(\eta) \right]
 \left[\bar{S}_{\bz\bv}(\eta) \bar{S}_{\bv\by}(\eta) - \bar{S}_{\bz\by}(\eta) \right]
  \Big\},
 \end{align}
which has a similar structure as Eq.(\ref{A5}). So we use the same scheme to simplify it. Using the property that the running coupling terms are invariant under $\bx\minus\bz \to \bz\minus\by$ and discarding the last term in the order of $\mathcal{O}(\abar^3)$, Eq.(\ref{B2}) can be reduced to
\begin{align}
\label{B11}
\frac{\partial \bar{S}_{\bx\by}(\eta)}{\partial \eta}& = \,
  \int
 \dif^2 \bz \, K^{\mathrm{rc}}(\bx,\by,\bz)
\left[\bar{S}_{\bx\bz}(\eta)\bar{S}_{\bz\by}(\eta)- \bar{S}_{\bx\by}(\eta)\right]
\nn
&-
2\int
	 \dif^2 \bz \,\dif^2 \bu K^{\mathrm{rc}}(\bx,\by,\bz)K^{\mathrm{rc}}(\bz,\by,\bu)
	\delta_{\bz\by;r}
	\bar{S}_{\bx\bz}(\eta)\left[\bar{S}_{\bz\bu}(\eta) \bar{S}_{\bu\by}(\eta) - \bar{S}_{\bz\by}(\eta) \right]
 .
 \end{align}

In order to keep consistence with the physics picture mentioned in the previous sections, we relabel the integral variables according to $\bu \leftrightarrow \bz$ for the second term in the right hand side Eq.(\ref{A11}), it becomes
   \begin{align}
\label{B12}
  \frac{\partial \bar{S}_{\bx\by}(\eta)}{\partial \eta}& = \,
  \int
 \dif^2 \bz \, K^{\mathrm{rc}}(\bx,\by,\bz)
\left[\bar{S}_{\bx\bz}(\eta)\bar{S}_{\bz\by}(\eta)- \bar{S}_{\bx\by}(\eta)\right]
\nn
&	-
2\int
	 \dif^2 \bz \,\dif^2 \bu K^{\mathrm{rc}}(\bx,\by,\bu)K^{\mathrm{rc}}(\bu,\by,\bz)
	\delta_{\bu\by;r}\
	\bar{S}_{\bx\bu}(\eta)
 \left[\bar{S}_{\bu\bz}(\eta) \bar{S}_{\bz\by}(\eta) - \bar{S}_{\bu\by}(\eta) \right]
 .
\end{align}
It is clear that the $\mathcal{O}(\abar^2)$ piece in the above equation is

   \begin{align}
\label{A13}
  &	-
2\int
	 \dif^2 \bz \,\dif^2 \bu K^{\mathrm{rc}}(\bx,\by,\bu)K^{\mathrm{rc}}(\bu,\by,\bz)
	\delta_{\bu\by;r}\
	\bar{S}_{\bx\bu}(\eta)
 \left[\bar{S}_{\bu\bz}(\eta) \bar{S}_{\bz\by}(\eta) - \bar{S}_{\bu\by}(\eta) \right]
 .
\end{align}


\end{document}